\documentclass[a4paper,11pt]{article}
\usepackage{jcappub}
\usepackage{lineno}

\title{\boldmath Diffuse Supernova Neutrino Background and Neutrino Non-Radiative Decay: a Bayesian Perspective}

\author[a,b]{Noah Roux}
\author[a,1]{and Maria Cristina Volpe\note{Corresponding author.}}
\affiliation[a]{CNRS, Université Paris Cité, Astroparticule et Cosmologie,\\ F-75013 Paris, France}
\affiliation[b]{Department of Physics, ETH Z\"urich,\\ 8093 Z\"urich, Switzerland}

\emailAdd{volpe@apc.in2p3.fr}

\abstract{Neutrinos being massive could undergo non-radiative decay, a property for which 
the diffuse supernova neutrino background has a unique sensitivity. We extend previous analyses to explore our ability to disentangle 
predictions for the diffuse supernova neutrino background in presence or absence of neutrino non-radiative two-body decay.
In a three-neutrino framework, we give predictions of the corresponding neutrino fluxes and the expected number of events in the Super-Kamiokande+Gadolinium, the Hyper-Kamiokande, the JUNO and the DUNE experiments. In our analysis, we employ supernova simulations from different groups and include 
current uncertainties from both the evolving core-collapse supernova rate and the fraction of failed supernovae. We perform the first Bayesian analysis 
to see our ability to disentangle the cases in presence and absence of neutrino decay. To this aim we combine the expected events in inverse beta-decay and the neutrino-argon detection channels. We also discuss neutrino-electron, neutrino-proton and of neutrino-oxygen scattering. Our investigation covers the different possible decay patterns for normal mass ordering, both strongly-hierarchical and quasi-degenerate as well as the inverted neutrino mass ordering.}

\usepackage[separate-uncertainty=true]{siunitx} 
\DeclareSIUnit\year{yr}
\DeclareSIUnit\erg{erg}
\DeclareSIUnit\parsec{pc}
 
\usepackage{multirow}
\usepackage{multicol}
\usepackage{tikz}
\usepackage[version=4]{mhchem} 
\usepackage{gensymb} 
\usepackage{subcaption} 

\begin{document}
\maketitle
\flushbottom

\section{Introduction} \label{sec:intro}

To date, SN1987A is the only event for which we observed neutrinos from the inner core of an exploding massive star~\cite{Hirata_1987, Bionta_1987, Alexeyev_1988}. 
Since, the time and energy spread of the 24 $\bar{\nu}_e$ detected constitute a unique laboratory for particle physics and astrophysics. In particular, it yielded the crucial confirmation that the total neutrino luminosity agrees with the expected emission of \(\SI{3e53}{\erg}\) of gravitational energy, in about ten seconds. While we expect a few galactic core-collapse supernovae per century, past core-collapse supernovae produced a relic neutrino flux, known as the diffuse supernova neutrino background (DSNB).

The DSNB encodes complementary information to a single supernova since its flux depends on the cosmological model, the evolving core-collapse supernova rate, the fraction of failed supernovae and binaries, neutrino flavor evolution in dense environments, and on physics beyond the Standard Model  (see the reviews~\cite{Ando_2004,Beacom_2010,Lunardini_2016,Volpe_2024}). The cosmological model usually considered is $\Lambda\text{CDM}$ for which anomalies and tensions (e.g. on the Hubble constant) are currently object of debate~\cite{Abdalla_2022}. Astrophysical uncertainties on the DSNB flux originate
from the evolving core-collapse supernova rate, the debated fraction of failed supernovae that determines the slope of the DSNB high energy tail, and the neutrino flux
from a given supernova. In particular, there is a disagreement by a factor of about 2 between the evolving core-collapse supernova rate from direct observations and the one deduced from the star formation history~\cite{Horiuchi_2011}. On the other hand, the fraction of failed supernovae is an important unknown. Direct observations yield the value $0.16^{+0.23}_{-0.12}$ at 90$\%$ confidence level~\cite{Neustadt_2021}, whereas detailed one-dimensional supernova simulations find the range \([0.18, 0.42]\)~\cite{Kresse_2021}.

The combined data analysis from the twenty-year data taking of the Super-Kamiokande (SK) experiment, SK-I to SK-IV, furnished the DSNB flux upper limit of 2.7~$\bar{\nu}_e$~\(\si{\per\centi\meter\squared\per\second}\) (\(E_{\nu} > \SI{17.3}{\mega\electronvolt}\), \(90\%\) C.L.)
and points to a $1.5\,\sigma$ excess over background (model-dependent analysis)~\cite{Abe_2021}. The SNO experiment obtained
the upper limit of 19~$\nu_e$~\(\si{\per\centi\meter\squared\per\second}\) (90$\%$ C.L.)~\cite{Aharmim_2020}; whereas from the analysis of SK-I data, ref.~\cite{Lunardini_2008} obtained the current bound for the $\nu_x$ flux, which could be improved by a factor of \(10^3\) thanks to dark matter detectors~\cite{Suliga_2022}. 

The Super-Kamiokande Collaboration just announced the results from SK-VI and SK-VII phases which include gadolinium (SK-Gd)\footnote{Gd concentrations are 0.01$\%$ and 
0.03$\%$ respectively.} to improve neutron tagging and background reduction, following the first suggestion by ref.~\cite{Beacom_2003}. According to the new model-dependent analysis, the significance of the excess over background has increased to $2.3\,\sigma$~\cite{Harada_2024}. 
These results might be an indication that the DSNB detection is imminent. While the SK-Gd experiment will run until 2027, the Hyper-Kamiokande (HK) detector~\cite{Abe_2018} should take over, with eight time more fiducial volume, the JUNO scintillator detector~\cite{An_2016, Abusleme_2022} should start soon, whereas the liquid argon DUNE detector~\cite{Acciarri_2016} is expected to start around 2030. Moreover novel detector technologies are under study, such as with the THEIA detector, 
that might increase the sensitivity and reduce backgrounds by combining the Cherenkov and scintillator technologies~\cite{Askins_2020}.

Astrophysical neutrinos have a unique sensitivity on non-standard neutrino properties, such as neutrino decay.
Tight bounds on neutrino radiative decay were obtained from SN1987A (see e.g.~\cite{Fogli_2004,rpp_2024}).
Neutrino non-radiative decay with lifetime-to-mass
ratio  $\tau/m$  in the range \([10^9, 10^{11}]\,\si{\second\per\electronvolt}\) can impact the DSNB, as pointed out by ref.~\cite{Ando_2004}. In a 3$\nu$ framework, ref.~\cite{Fogli_2004} investigated the effect considering the two possible neutrino mass orderings and different mass patterns. Their results showed that neutrino non-radiative decay influences the DSNB rates for inverse beta-decay in a significant way. Using an effective \(2\nu\) approach, ref.~\cite{DeGouvea_2020_bis} studied the prospects in SK, HK, DUNE, JUNO and THEIA, considering normal mass ordering and a strongly hierarchical mass pattern only. These works assumed for the supernova fluxes pinched Fermi-Dirac distributions, neglecting any progenitor mass dependence. With an effective 2$\nu$ flavor approach, ref.~\cite{Tabrizi_2021} considered both supernova neutrino fluxes from two-dimensional simulations for different progenitors and Fermi-Dirac distributions, and explored the possibility to disentangle the DNSB rates in absence and in presence of neutrino decay, for normal mass ordering and a strongly hierarchical mass pattern. By combining detection channels in JUNO, HK and DUNE, they concluded that discriminating them was possible (to some degree), in particular for \(\tau/m=10^{9}\,\si{\second\per\electronvolt}\), through the inclusion of neutrino-proton scattering. 

More recently, ref.~\cite{MartinezMirave_2024} investigated non-radiative decay to invisible neutrinos, combining HK, JUNO and DUNE and obtained expected lower bounds \(\tau_1 / m_1 > 4.2~ (4.4) \times10^8\,\si{\second\per\electronvolt}\) (at $90\%$ C.L.) for normal (inverted) ordering. Finally, ref.~\cite{IvanezBallesteros_2023} investigated neutrino non-radiative decay in the 2$\nu$ and 3$\nu$ frameworks including both a progenitor mass dependence and the astrophysical uncertainty from the evolving core-collapse supernova rate. Their results showed that for inverted mass ordering, neutrino decay could significantly suppress the DSNB rates; whereas for normal mass ordering (and any mass pattern), the results confirmed the presence of important degeneracies between the rates that include and those that do not include neutrino decay.

In this work we present an investigation of the DSNB in a 3$\nu$ framework including neutrino non-radiative two-body decay.  
Our study implements the progenitor mass dependence, the fraction of failed supernovae and astrophysical uncertainties from the evolving core-collapse supernova rate
and, to some extent, the neutrino fluxes from a individual supernovae. For the latter, one-dimensional simulations from the Garching and Nakazato's groups were considered. 
We give predictions for the DSNB fluxes as well as the number of events for the running SK-Gd, the upcoming JUNO and the near-future HK and DUNE experiments,
considering as detection channels inverse beta-decay and neutrino scattering on electrons, protons, argon and oxygen nuclei. 
We perform the first Bayesian analysis in order to assess our ability to disentangle the DSNB rates in presence of decay, from those in absence,
for the cases of normal mass ordering and strongly hierarchical or quasi-degenerate mass patterns as well as of inverted mass ordering. 
Finally, we provide Bayes factors when uncertainties for the signal and backgrounds are included, in either a conservative or 
an optimistic scenario. For completeness, we present at the end the Bayes factors one obtains if neutrinos undergo invisible neutrino decay.

The manuscript is structured as follows. Section~\ref{sec:non_radiative_two_body_decay} introduces the process of neutrino non-radiative decay and section~\ref{sec:DSNB} 
describes the DSNB flux and its cosmological and astrophysical ingredients. In particular, the section describes the dependence on the cosmological model, on the evolving core-collapse supernova rate, the progenitor mass dependence and supernova simulations. Section~\ref{sec:DSNB_with_decay} details the influence of non-radiative decay on the DSNB flux, via the solution of the neutrino kinetic equations, and discusses the different possible decay patterns. The Bayesian framework is presented in section~\ref{sec:analysis}. 
Section~\ref{sec:numerical_fluxes_and_rates} focuses on the numerical results with and without decay, for the DSNB fluxes and rates in SK-Gd, HK, JUNO and DUNE. The outcome of the Bayesian analysis is provided in section~\ref{sec:analysis_results}. Finally section~\ref{sec:conclusions} is a conclusion.

\section{Neutrino non-radiative two-body decay} \label{sec:non_radiative_two_body_decay}

Neutrinos being massive, they could decay. Indeed a heavier neutrino mass eigenstate $\nu_h$ can decay into a lighter (anti)neutrino $\nu_l$ ($\bar{\nu}_l$) and a massless scalar or pseudoscalar particle such as a Majoron~\cite{Chikashige_1981}, i.e.
\begin{equation}\label{eq:decay}
    \nu_h \rightarrow \nu_l  + \varphi ~~~~{\rm or} ~~~~ \nu_h \rightarrow \bar{\nu}_l + \varphi\,.
\end{equation}
Here we consider that decay occurs in vacuum and that the decaying eigenstates coincide with the mass eigenstates (for a discussion on the effects of a possible mismatch see refs.~\cite{Berryman_2015,Chattopadhyay_2022}). Note that neutrino non-radiative decay in matter into a massless (pseudo)scalar was considered in refs.~\cite{Kachelriess_2000, Farzan_2003,IvanezBallesteros_2024}. We do not consider here specific models for the decay to keep our discussion and the conclusions general. 

The decay eq.~\eqref{eq:decay} depends on the neutrino nature. If neutrinos are Majorana particles, the scalar does not have definite lepton number (see ref.~\cite{Kim_1990}) and a dimension six operator is required for the decay~\cite{DeGouvea_2020}
\begin{equation} \label{eq:phi}
    \nu_{h,L}\rightarrow\nu_{l,L}+\varphi \hspace{0.7cm}\text{and}\hspace{0.7cm} \nu_{h,L}\rightarrow\nu_{l,R}+\varphi \,.
\end{equation}
If neutrinos are Dirac particles, the decay into a lepton-number-zero scalar $\varphi_0$ requires a dimension 5 operator at lowest order, allowing for decays of the form
\begin{equation} \label{eq:phi0_decay}
    \nu_{h,L}\rightarrow\nu_{l,L}+\varphi_0 \hspace{0.7cm}\text{and}\hspace{0.7cm} \nu_{h,L}\rightarrow\nu_{l,R}+\varphi_0\,,
\end{equation}
whereas dimension 4 or 6 operators are needed for decays into a lepton-number-two scalar~$\varphi_2$ 
\begin{equation} \label{eq:phi2_decay}
    \nu_{h,L}\rightarrow\bar\nu_{l,L}+\varphi_2 \hspace{0.7cm}\text{and}\hspace{0.7cm} \nu_{h,L}\rightarrow\bar\nu_{l,R}+\varphi_2\,.
\end{equation}

In the laboratory frame, the neutrino lifetime for processes \eqref{eq:decay} is related to the decay rate via
\begin{equation}\label{eq:tau}
    \Gamma_{\nu_h} =  \frac{m_h}{E_{\nu}}\tau^{-1}_{\nu_h} = \frac{m_h}{E_{\nu}}\sum_{m_h > m_l} [ \tilde{\Gamma}(\nu_h \rightarrow \nu_l ) +\tilde{\Gamma}(\nu_h \rightarrow \bar{\nu}_l) ] \,,
\end{equation}
where $\tau_{\nu_h} $ and $\tilde{\Gamma}$ denote the lifetime and the decay rate in the rest frame, respectively, $E_{\nu}/m_h$ being the boost factor with $E_{\nu}$ the neutrino energy. For the process of neutrino decay, bounds are usually given for the lifetime-to-mass ratio $\tau/m$, since the absolute neutrino mass remains unknown (the current upper limit is $ \langle m_{\nu_e} \rangle < \SI{0.45}{\electronvolt}$ at 90 $\%$ C.L. from the KATRIN experiment~\cite{Katrin_2024}). 

\section{The DSNB flux in absence of neutrino decay} \label{sec:DSNB}
The neutrino diffuse background from past core-collapse supernovae depends on astrophysical as well as on cosmological inputs (see~\cite{Ando_2004, Beacom_2010,Lunardini_2016,Mathews_2019,Volpe_2024} for reviews).
On Earth the DSNB flux reads 
\begin{equation} \label{eq:flux}
    \begin{aligned}
        \phi_{\nu_i}(E_\nu) = c \int_0^{\infty}  \mathrm{d}z (1+z) \left| \frac{\mathrm{d}t_\mathrm{c}}{\mathrm{d}z} \right| &\left[ \int_\Omega \mathrm{d}M R_{\text{SN}}(z, M) Y_{\nu_i, {\rm NS}}(E_\nu', M) \right. \\
        &\left.\quad+ \int_\Sigma \mathrm{d}M R_{\text{SN}}(z, M) Y_{\nu_i, {\rm BH}}(E_\nu', M) \right]\,,
    \end{aligned}
\end{equation}
where $c$ is the speed of light, $z$ the cosmological redshift, and $E'_{\nu}=(1+z) E_{\nu}$ the redshifted neutrino energy. The quantity $\left| {\rm d}t_{\rm c}/{{\rm d } z} \right|$
is the cosmic time. We consider progenitor masses in the range $[8, 125]\,M_\odot$ ($M_\odot$ being the solar mass) and account for the fact that the collapse can produce
either a neutron star (NS) or a black hole (BH), $\Omega$ and $\Sigma$ being the corresponding progenitor mass ranges. Indeed, as pointed out in ref.~\cite{Lunardini_2009}, the contribution from failed supernovae, can be important although sub-leading since the compression of baryonic matter produces a hardening of the relic fluxes.
As for the maximum redshift, we take $z = 5$ as usually done, the most important contribution to the DSNB flux coming from $z \in [0,3]$. 

Let us now describe the three main ingredients in eq.~(\ref{eq:flux}), i.e. the physical inputs from cosmology, astrophysics, and the neutrino yields.
Cosmology impacts the DSNB flux through the cosmic time. Assuming the $\Lambda {\rm CDM}$ model, the expansion history of the universe is given by
\begin{equation}
    \left| \frac{{\rm d}z}{{\rm d}t_{\rm c}} \right| = H_0(1+z)\sqrt{\Omega_{\Lambda} + (1+z)^3 \Omega_m} \,,
\end{equation}
with $H_0= \SI{70}{\kilo\meter\per\second\per\mega\parsec}$ the Hubble constant, whereas $\Omega_{\Lambda} = 0.7$ and $\Omega_m = 0.3$ are the dark energy and matter cosmic energy densities, respectively. Note that the value of $H_0$ is currently debated~\cite{Abdalla_2022}. 

\begin{figure}[htbp]
    \centering
    \includegraphics[trim=0 15 0 5, clip, width=0.5\textwidth]{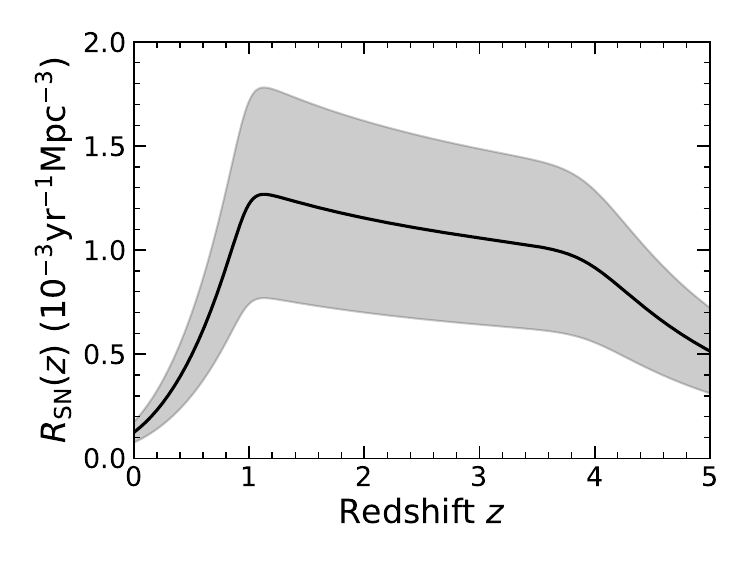}
    \caption{Evolving core-collapse supernova rate as a function of cosmological redshift $z$ following the broken power-law parametrization for the evolving star-formation rate of ref.~\cite{Yuksel_2008}. The band shows the astrophysical uncertainty coming from the local rate $R_{\rm SN}(0)$.}
    \label{fig:R_SN}
\end{figure}

Concerning the evolving core-collapse supernova rate\footnote{This corresponds to the number of supernovae occurring per unit time per unit comoving volume.} $R_{\rm{SN}}(z, M)$, it is related to the star-formation rate history \(\dot\rho_*(z)\) and the initial mass function (IMF) $\phi(M)$ via\footnote{Note that, while lower bounds of the denominator integral different from \(0.5M_{\odot}\) are also used in the literature (see e.g. ref.~\cite{Nakazato_2015}), the impact on $R_{\rm SN} (z,M)$ can be mitigated through the normalization of the star-formation rate history~\cite{Volpe_2024}.}
\begin{equation}\label{eq:rsnz}
    R_{\rm SN}(z,M) = \frac{\dot{\rho}_*(z) \phi(M)}{\int_{0.5M_\odot}^{125M_\odot} \phi(M) M \, dM}\,.
\end{equation}
We assume the IMF introduced by Salpeter~\cite{Salpeter_1955} (see e.g.~\cite{Ziegler_2022} for a discussion on its universality)
\begin{equation}
    \phi(M) \sim M^\chi ~~~ {\rm with }~~~ \chi=-2.35 \,, 
\end{equation}
with the quantity $\phi(M){\rm d}M$ being the number of stars in the mass interval $[M, M+{\rm d}M]$.
As for the star formation rate history, we employ the following broken power-law parametrization~\cite{Yuksel_2008} 
\begin{equation}
\dot{\rho}_*(z) = \dot{\rho}_0 \left[ (1 + z)^{\alpha \eta} + \left( \frac{1 + z}{B} \right)^{\beta \eta} + \left( \frac{1 + z}{C} \right)^{\gamma \eta} \right]^{1/\eta}
\end{equation}
with the logarithmic slopes $\alpha = 3.4$, $\beta = -0.3$, $\gamma=-3.5$ at low, intermediate and high redshift, respectively. The parameters $B=5000$ and $C=9$ define the redshift breaks, and $\eta=-10$ is a smoothing parameter. The local star formation rate history \(\dot\rho_0\) was adjusted in the calculations to obtain the desired local core-collapse supernova rate\footnote{See table~I in ref.~\cite{IvanezBallesteros_2023} for explicit values.}. Figure~\ref{fig:R_SN} shows the evolving core-collapse supernova rate $R_{\rm SN}(z)$ eq.~\eqref{eq:rsnz} used, along with the variability induced by the uncertainty on the local rate (integrated over progenitor masses) 
\begin{equation}
R_{\rm SN}(0) = \int_{8M_\odot}^{125M_\odot} R_\text{SN}(0, M) \, {\rm d}M \\
= \SI{1.25\pm 0.5 e-4}{\per\year\per\cubic\mega\parsec}\,.
\end{equation}

\subsection{Scenarios for the neutrino yields at given supernovae}
One of the key ingredients in the DSNB flux are the neutrino yields $Y_{\nu, {\rm NS}}$ and $Y_{\nu, {\rm BH}}$ for NS- and BH-forming supernovae, respectively.
Such yields depend on the progenitor, the neutrino spectra at the neutrinosphere and the flavor mechanisms that neutrinos undergo while traversing the dense medium up to the stellar surface. To determine the two contributions to the DSNB flux eq.~\eqref{eq:flux} we used scenarios based on two different sets of detailed supernova simulations, from the Garching group and from Nazakato's groups. The scenarios also differ in the fraction of BH-forming supernovae defined as
\begin{equation}
    f_{\rm BH} = \frac{\int_{\Sigma} {\rm d}M\phi(M)}{\int_{8M_\odot}^{125M_\odot} {\rm d}M}
\end{equation}
which parametrizes the contribution from failed supernovae to the DSNB flux.

\paragraph{Scenarios based on Nakazato simulations}
The first set of simulations, referred to hereafter as ``Nakazato''\footnote{Simulation data are available on \url{http://asphwww.ph.noda.tus.ac.jp/snn/}.}~\cite{Nakazato_2013}, consists of spherically symmetric models of progenitors with metallicities $Z=0.02$ or $Z=0.004$ and shock revival time $t_{\rm rev}$=100, 200, or \(\SI{300}{\milli\second}\) for the NS-forming progenitors. Whenever necessary, a log-linear interpolation or extrapolation of the simulated numerical yields as a function of the neutrino energy was applied.

\begin{figure}[htbp]
    \centering
    \includegraphics[trim=0 0 0 30, clip, width=0.99\textwidth]{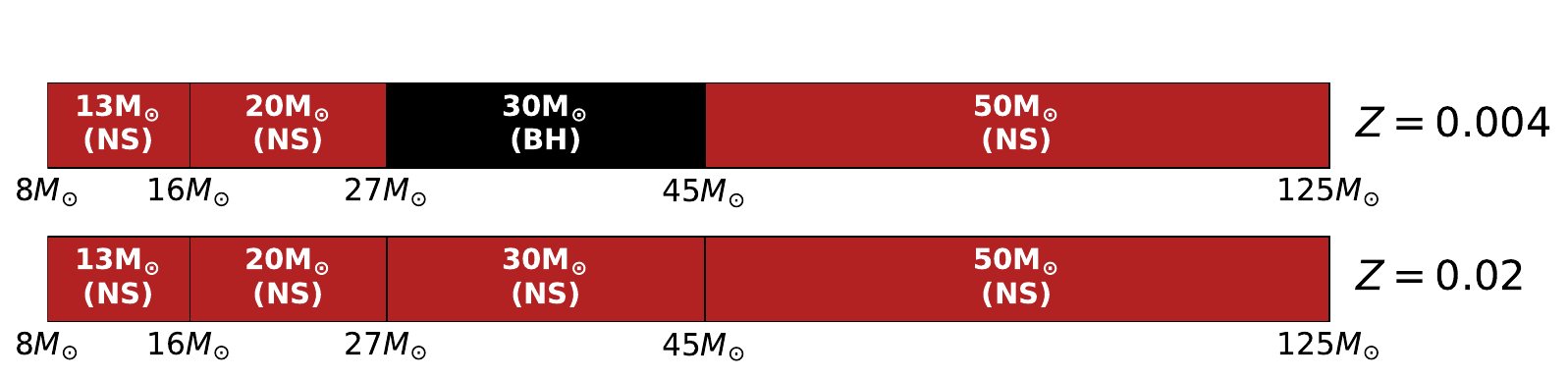}
    \caption{The figure shows the progenitor mass templates and corresponding ranges used for the Nakazato simulations, for each of the metallicities \(Z=0.004\) (top) and \(Z=0.02\) (bottom). The progenitor masses and their fate are indicated in the boxes, while the bounds underneath indicate the progenitor mass range for which the progenitor (in the box) is used as a reference. Such ranges define the \(\Omega\) and \(\Sigma\) sets contributing to the DSNB flux, either from NS- or for BH-forming supernovae.}
    \label{fig:nakazato_templates}
\end{figure}

The equation of states (EoS) considered is the one by Shen\footnote{Note that for example ref.~\cite{Nakazato_2021} studied the impact of the nuclear EoS on the DSNB flux (see also~\cite{Volpe_2024}).}~\cite{Shen_1998, Shen_1998_bis}. For both metallicities, the simulations include four progenitors with respective masses $13M_\odot$, $20M_\odot$, $30M_\odot$, and $50M_\odot$. All progenitors, except the one with mass $30M_\odot$ and metallicity $Z=0.004$, form a neutron star. Note that the simulations for the NS-forming supernovae extend up to \(\SI{20}{\second}\) after the core bounce. The templates used to associate a given progenitor mass to one of the simulated progenitors are shown in figure~\ref{fig:nakazato_templates}. 
For the fraction of failed supernovae, the simulations with solar metallicity $Z=0.02$, we take $ f_{\rm BH} = 0$; while for the ones with the low metallicity $Z = 0.004$ 
we take $f_{\rm BH} = 0.14$. 

\paragraph{Scenarios based on Garching simulations}
The second set of scenarios exploits one-dimensional simulations by the Garching group~\cite{Priya_2017, Hudepohl_2013} using the Lattimer-Swesty equation of state with compressibility parameter \(K=220\) (LS220)~\cite{Lattimer_1991}. It includes simulations of three solar-metallicity NS-forming progenitors with masses \(11.2M_\odot\), \(25M_\odot\) and \(27M_\odot\), as well as two solar-metallicity BH-forming progenitors with masses \(25M_\odot\) and \(40M_\odot\). 

Following refs.~\cite{Priya_2017, Moeller_2018, IvanezBallesteros_2023}, we employed three distinct scenarios for the fraction of BHs, i.e. \(f_\text{BH}=0.09, 0.21\text{~and~}0.41\). The smallest value is taken for comparison with the results presented in refs.~\cite{Priya_2017, IvanezBallesteros_2023}. The other two values are in agreement with the outcome of the detailed simulations in ref.~\cite{Kresse_2021}. 
\begin{figure}[htb]
    \centering
    \includegraphics[trim=0 0 0 30, clip, width=0.99\textwidth]{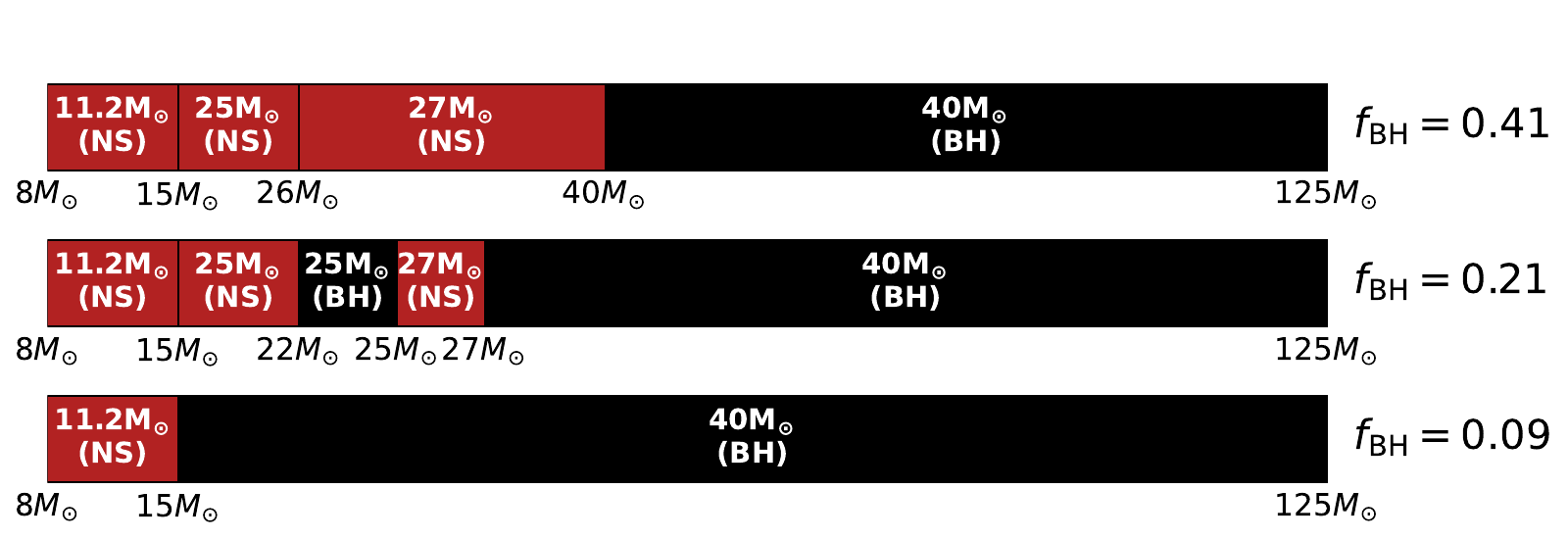}
    \caption{The figure shows the progenitor mass templates and corresponding ranges used for the Garching simulations with fraction of failed supernovae 
    $f_{\rm BH}=0.41$  (top), $f_{\rm BH}=0.21$ (middle), and $f_{\rm BH}=0.09$ (bottom). The case with $f_{\rm BH} = 0.21$ is our "reference" scenario. The progenitor masses and their fate are indicated in the boxes, while the bounds underneath indicate the progenitor mass range for which the progenitor (in the box) is used as a reference. Such ranges define the \(\Omega\) and \(\Sigma\) sets contributing to the DSNB flux, either from NS- or for BH-forming supernovae. }
    \label{fig:garching_templates}
\end{figure}

While in the Nakazato case the neutrino yields for each progenitor are obtained through interpolation of the numerical results, 
the normalized neutrino spectra from the Garching simulations follow a power-law distribution 
\begin{equation}\label{eq:power_law}
    \varphi_{\nu}^{0}(E_{\nu}) = \frac{(\alpha + 1)^{\alpha + 1}}{\langle E_{\nu} \rangle \Gamma(\alpha + 1)} \left( \frac{E_{\nu}}{\langle E_{\nu} \rangle} \right)^{\alpha} \exp \left( - \frac{(1 + \alpha) E_{\nu}}{\langle E_{\nu} \rangle} \right)\,,
\end{equation}
where \(\langle E_\nu\rangle\) is the mean neutrino energy and $\alpha$ the pinching parameter. 
The latter can be written in terms of the mean energy squared as
\begin{equation}
    \alpha = \frac{\langle E_{\nu}^{2} \rangle - 2 \langle E_{\nu} \rangle^{2}}{\langle E_{\nu} \rangle^{2} - \langle E_{\nu}^{2} \rangle}\,.
\end{equation}
The neutrino yields at the neutrinosphere are then obtained by scaling the spectra with the ratio between the total neutrino luminosity \(L_\nu\) and the mean neutrino energy:
\begin{equation}\label{eq:yields}
    Y^0_{\nu}(E_\nu, M) = \frac{L_{\nu}}{\langle E_{\nu} \rangle} \varphi_{\nu}^{0}(E_{\nu})\,,
\end{equation}
The three parameters \(\langle E_\nu\rangle\), \(\langle E^2_\nu\rangle\) and \(L_\nu\) used for each progenitor mass $M$ can be found in appendix~A of ref.~\cite{IvanezBallesteros_2023}.

In both Garching and Nakazato scenarios we consider that neutrinos propagating from the neutrinosphere to the star surface are subject to flavor transformation phenomena. Here we consider only the established MSW effect~\cite{Wolfenstein_1978, Mikheev_1986}, due to neutrino-matter interactions.
We do not consider flavor mechanisms due to the \(\nu\nu\) interactions, to turbulence, and to shock waves that can also modify the neutrino spectra at a given supernova~(see ref.~\cite{Volpe_2024} for a review). 

Because of the MSW effect, the (anti)neutrino yields at the star surface are affected by spectral modifications due to the high- and low-resonances~\cite{Dighe_2000}. As a result, in the normal mass ordering (NO), i.e. \(\Delta m^2_{31}>0\), they are given by 
\begin{equation}\label{eq:MSW_NO}
    \begin{array}{lll}
    Y_{\nu_1}=Y^0_{\nu_x}\,, & \hspace{0.2cm} Y_{\nu_2}=Y^0_{\nu_x}\,, & \hspace{0.2cm}Y_{\nu_3}=Y^0_{\nu_e} \\[0.18cm]
    Y_{\bar\nu_1}=Y^0_{\bar\nu_e}\,, & \hspace{0.2cm} Y_{\bar\nu_2}=Y^0_{\nu_x}\,, & \hspace{0.2cm} Y_{\bar\nu_3}=Y^0_{\nu_x}
    \end{array} \ , 
\end{equation}
while in the inverted mass ordering (IO), i.e. \(\Delta m^2_{31}<0\) they read
\begin{equation}\label{eq:MSW_IO}
    \begin{array}{lll}
    Y_{\nu_1}=Y^0_{\nu_x}\,, & \hspace{0.2cm} Y_{\nu_2}=Y^0_{\nu_e}\,, & \hspace{0.2cm}Y_{\nu_3}=Y^0_{\nu_x}\,  \\[0.18cm]
    Y_{\bar\nu_1}=Y^0_{\nu_x}\,, & \hspace{0.2cm} Y_{\bar\nu_2}=Y^0_{\nu_x}\,, & \hspace{0.2cm} Y_{\bar\nu_3}=Y^0_{\bar\nu_e}
    \end{array} \ ,
\end{equation}
where $\nu_i$ with $i=1,2,3$ refers to the neutrino mass eigenstate. For the non-electron flavors \(\nu_x\), we make the usual assumption \(Y^0_{\nu_\mu}=Y^0_{\bar\nu_\mu}=Y^0_{\nu_\tau}=Y^0_{\bar\nu_\tau}=Y^0_{\nu_x}\).

\section{The DSNB in presence of neutrino non-radiative decay} \label{sec:DSNB_with_decay}
We now discuss how the DSNB flux \eqref{eq:flux} is modified in presence of neutrino decay. To this aim we consider neutrino kinetic equations including neutrino decay
as done in ref.~\cite{Fogli_2004}. Then, we present the possible decaying schemes for inverted neutrino mass ordering or for normal mass ordering with either a strongly hierarchical or a quasi-degenerate neutrino mass pattern. 

\subsection{Kinetic equations with neutrino decay}
The kinetic equations for ultrarelativistic neutrinos take the generic form~\cite{Fogli_2004}
\begin{equation}\label{eq:qke}
    {\cal L} [n_{\nu_k}(E_{\nu}, t, M) ] =  {\cal C} [n_{\nu_k}(E_{\nu}, t, M) ]\,,
\end{equation}
where ${\cal L}$ is the Liouville and ${\cal C}$ the collision operator, and $n_{\nu_k}(E_{\nu},t,M)$ is the relic number density of the $\nu_k$ mass eigenstates (per unit energy, comoving volume and progenitor mass) at time $t$ and progenitor mass $M$.
More explicitly, the Liouville operator reads
\begin{equation}
    {\cal L}  [n_{\nu_k}(E_{\nu}, t, M) ] =  \left[ \partial_t - H(t) E_{\nu} \partial_{E_\nu} - H(t) \right]  n_{\nu_k}(E_{\nu}, t, M)\,,
\end{equation}
with $H(t)$ the Hubble parameter, whereas the collision term is given by
\begin{equation}\label{eq:collision}
    {\cal C}  [ n_{\nu_k}(E_{\nu}, t, M) ] =  R_{\rm SN} (t, M) Y_{\nu_k} (E_{\nu}, M) + \sum_{m_i > m_k} q_{ik} (E_{\nu},t, M) - \Gamma_{\nu_k} n_{\nu_k}(E_{\nu}, t, M)\,,
\end{equation}
where
\begin{equation}\label{eq:qji}
q_{ik} (E_{\nu},t, M) = \int_{E_{\nu}}^{\infty} \mathrm{d}\tilde E_{\nu} n_{\nu_i}(\tilde E_{\nu}, t, M) \Gamma_{\nu_i \rightarrow \nu_k} \psi_{ik}(\tilde E_{\nu}, E_{\nu})\,, 
\end{equation}
with \(\psi_{ik}(\tilde E_{\nu}, E_{\nu})\) the neutrino energy spectra after decay. While the first contribution to the collision term~(\ref{eq:collision}) is the usual one from core-collapse supernovae, the second and third contributions are a source and a sink term due to neutrino decay. The source term, of course, vanishes for the heaviest (anti)neutrino, and the sink term vanishes for the lightest one.

By performing the change of variables \((t, E_\nu)\rightarrow (z, E'_\nu)\) where $E_{\nu} = E'_{\nu}(1 + z)$, one can readily check that~\cite{Fogli_2004}
\begin{equation}\label{eq:solution}
    \begin{aligned}
        n_{\nu_k} (E_{\nu},z, M) = \frac{1}{1+z} \int_z^{z_\text{max}} \frac{\mathrm{d} z'}{H(z')} &\bigg[R_\mathrm{SN}(z', M) Y_{\nu_k} \left(E_{\nu} \frac{1 + z'}{1 + z}, M\right)\\
        &\quad+ \sum_{ m_i > m_k} q_{ik} \left(E_{\nu} \frac{1 + z'}{1 + z}, z', M\right)  \bigg]  e^{- \Gamma_{\nu_k}[\chi(z') - \chi(z)](1 + z)}
    \end{aligned}
\end{equation}
solves the kinetic equations~(\ref{eq:qke}), with the auxiliary function
\begin{equation}
    \chi(z) = \int_0^z \mathrm{d}z' H^{-1}(z')(1+z')^{-2} \ .
\end{equation}
The no-decay case is obviously recovered by taking vanishing decay rates. 

\subsection{Neutrino decay schemes}
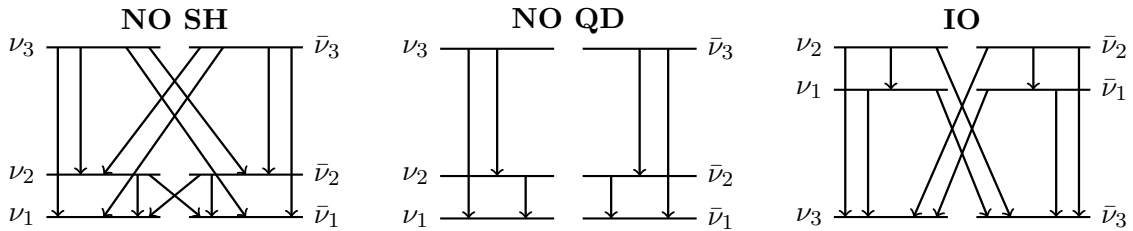
\begin{figure}[htbp]
    \centering
    \begin{minipage}{0.33\textwidth}
        \centering
        \begin{tikzpicture}[scale=0.75] 

            \node at (2.25,3.5) {\textbf{NO SH}};

            \draw[thick] (0,0) node[anchor=east] {$\nu_1$} -- (2,0) ;
            \draw[thick] (0,0.75) node[anchor=east] {$\nu_2$} -- (2,0.75) ;
            \draw[thick] (0,3) node[anchor=east] {$\nu_3$} -- (2,3) ;
            \draw[->, line width=0.3mm] (0.2,3) -- (0.2,0.0);
            \draw[->, line width=0.3mm] (0.6,3) -- (0.6,0.75);
            \draw[->, line width=0.3mm] (1.6,0.75) -- (1.6,0.0);
            \draw[->, line width=0.3mm] (3.1,3.0) -- (1.0,0.0);
            \draw[->, line width=0.3mm] (2.7,3.0) -- (1.0,0.75);
            \draw[->, line width=0.3mm] (2.7,0.75) -- (1.8,0.01);

            \draw[thick] (2.5,0) -- (4.5,0) node[anchor=west] {$\bar{\nu}_1$};
            \draw[thick] (2.5,0.75) -- (4.5,0.75) node[anchor=west] {$\bar{\nu}_2$};
            \draw[thick] (2.5,3) -- (4.5,3) node[anchor=west] {$\bar{\nu}_3$};
            \draw[->, line width=0.3mm] (4.3,3) -- (4.3,0.0);
            \draw[->, line width=0.3mm] (3.9,3) -- (3.9,0.75);
            \draw[->, line width=0.3mm] (2.9,0.75) -- (2.9,0.0);
            \draw[->, line width=0.3mm] (1.4,3.0) -- (3.5,0.0);
            \draw[->, line width=0.3mm] (1.8,3.0) -- (3.5,0.75);
            \draw[->, line width=0.3mm] (1.8,0.75) -- (2.7,0.01);
        \end{tikzpicture}
    \end{minipage}
    \hfill
    \hspace{-1cm}
    \begin{minipage}{0.33\textwidth}
        \centering
        \begin{tikzpicture}[scale=0.75] 

            \node at (2.25,3.5) {\textbf{NO QD}};

            \draw[thick] (0,0) node[anchor=east] {$\nu_1$} -- (2,0) ;
            \draw[thick] (0,0.75) node[anchor=east] {$\nu_2$} -- (2,0.75) ;
            \draw[thick] (0,3) node[anchor=east] {$\nu_3$} -- (2,3) ;
            \draw[->, line width=0.3mm] (0.5,3) -- (0.5,0.0);
            \draw[->, line width=0.3mm] (1,3) -- (1,0.75);
            \draw[->, line width=0.3mm] (1.5,0.75) -- (1.5,0.0);

            \draw[thick] (2.5,0) -- (4.5,0) node[anchor=west] {$\bar{\nu}_1$};
            \draw[thick] (2.5,0.75) -- (4.5,0.75) node[anchor=west] {$\bar{\nu}_2$};
            \draw[thick] (2.5,3) -- (4.5,3) node[anchor=west] {$\bar{\nu}_3$};
            \draw[->, line width=0.3mm] (4.0,3) -- (4.0,0.0);
            \draw[->, line width=0.3mm] (3.5,3) -- (3.5,0.75);
            \draw[->, line width=0.3mm] (3.0,0.75) -- (3.0,0.0);
        \end{tikzpicture}
    \end{minipage}
    \hfill
    \hspace{-1cm}
    \begin{minipage}{0.33\textwidth}
        \centering
        \begin{tikzpicture}[scale=0.75] 
            \node at (2.25,3.5) {\textbf{IO}};

            \draw[thick] (0,0) node[anchor=east] {$\nu_3$} -- (2,0) ;
            \draw[thick] (0,2.25) node[anchor=east] {$\nu_1$} -- (2,2.25) ;
            \draw[thick] (0,3) node[anchor=east] {$\nu_2$} -- (2,3) ;
            \draw[->, line width=0.3mm] (0.2,3) -- (0.2,0.0);
            \draw[->, line width=0.3mm] (1.0,3) -- (1.0,2.25);
            \draw[->, line width=0.3mm] (0.6,2.25) -- (0.6,0.0);
            \draw[->, line width=0.3mm] (2.7,3.0) -- (1.4,0.0);
            \draw[->, line width=0.3mm] (2.7,2.25) -- (1.8,0.0);

            \draw[thick] (2.5,0) -- (4.5,0) node[anchor=west] {$\bar{\nu}_3$};
            \draw[thick] (2.5,2.25) -- (4.5,2.25) node[anchor=west] {$\bar{\nu}_1$};
            \draw[thick] (2.5,3) -- (4.5,3) node[anchor=west] {$\bar{\nu}_2$};
            \draw[->, line width=0.3mm] (4.3,3) -- (4.3,0.0);
            \draw[->, line width=0.3mm] (3.5,3) -- (3.5,2.25);
            \draw[->, line width=0.3mm] (3.9,2.25) -- (3.9,0.0);
            \draw[->, line width=0.3mm] (1.8,3.0) -- (3.1,0.0);
            \draw[->, line width=0.3mm] (1.8,2.25) -- (2.7,0.0);
        \end{tikzpicture}
    \end{minipage}
    \caption{Mass patterns for 3$\nu$ flavors for the cases of NO SH (left), NO QD (center) and of IO (right). We make a democratic assumption in which the branching ratios for NO SH are \(1/4\) for \(\nu_3\) and \(\bar\nu_3\), and \(1/2\) for \(\nu_2\) and \(\bar\nu_2\). For NO QD, they are \(1/2\) for \(\nu_3\) and \(\bar\nu_3\), and  \(1\) for \(\nu_2\) and \(\bar\nu_2\). For IO, the \(\nu_2\) and \(\bar\nu_2\) branching ratios are equal to 1/3 and those for \(\nu_1\) and \(\bar\nu_1\) are equal to 1/2. We assume each of the decaying eigenstates (assumed to be equal to the mass eigenstates) to have the same lifetime-to-mass ratio $\tau/m$ to have only one free parameter.}
    \label{fig:decay_patterns}
\end{figure}

In order to determine the DSNB fluxes, one needs to define the mass patterns for neutrino decay.
Here we shall consider the cases of both NO and IO, since the neutrino mass ordering has not been determined yet (normal ordering is currently favored at the \(2.5\,\sigma\) level~\cite{Capozzi_2021}). We consider two extreme conditions for the mass patterns, namely
\begin{itemize}
    \item the strongly hierarchical (SH) mass pattern, if \(m_h - m_l \gg m_l \simeq 0\) and
    \item the quasidegenerate (QD) mass pattern, if \(m_h \simeq m_l \gg m_h - m_l\)
\end{itemize}
for $m_l<m_h$.
The decay mass patterns for the cases of NO SH, NO QD and IO are shown in figure~\ref{fig:decay_patterns} in the $3 \nu$ framework\footnote{Note that DSNB predictions in the \(2\nu\) effective framework can be obtained by ``freezing'' the decays of one of the mass eigenstates and taking the branching ratios $B_{\nu_h\rightarrow\nu_l}=B_{\nu_h\rightarrow\bar\nu_l}=1/2$ (SH case), or \(B_{\nu_h\rightarrow\nu_l}=1\) and \(B_{\nu_h\rightarrow\bar\nu_l}=0\) (QD case).}. The case of IO comprises 
a QD $m_2$ and $m_1$ subsystem, whereas both theses states are SH with respect to $m_3$. Note that current cosmological bounds on the sum of neutrino masses~\cite{rpp_2024}
tend to favor the SH mass pattern\footnote{However note that cosmological bounds assume stable neutrinos and that they can be relaxed in scenarios of neutrino decay~\cite{Abellan_2022}.}.

Let us now describe the two ingredients required to characterize a decay scenario \eqref{eq:qji}. First, one needs to specify the branching ratios
\begin{equation}\label{eq:BR}
    B_{\nu_h\rightarrow\nu_{l}} = \Gamma_{\nu_h \rightarrow \nu_l}/\Gamma_{\nu_h}\,,
\end{equation}
and similarly for $\nu_h \rightarrow \bar{\nu}_l + \phi$ or the decay of antineutrinos. We make a democratic assumption for the branching ratios (explicit values are given in the caption of figure~\ref{fig:decay_patterns}) and assume equal lifetime-to-mass ratios for the decaying eigenstates to reduce the number of free parameters to a single $\tau/m$. Second, one needs to know the energy spectra. In the QD case, which can only exhibit helicity-conserving decays, the energy spectrum takes the form~\cite{Fogli_2004}
\begin{equation} \label{eq:qd_spectrum}
    \psi_{hl} (E_{\nu_h}, E_{\nu_l}) = \delta(E_{\nu_h} -E_{\nu_l})\,. 
\end{equation}
The SH case includes both helicity-conserving (h.c.) and helicity-flipping (h.f) decays, with respective spectra given by
\begin{equation} \label{eq:sh_spectrum}
    \psi_{hl,{\rm h.c.}} (E_{\nu_h}, E_{\nu_l}) = \frac{2 E_{\nu_l}}{E_{\nu_h}^2}~~~~~\text{and}~~~~~\psi_{hl,{\rm h.f.}} (E_{\nu_h}, E_{\nu_l}) =  \frac{2}{E_{\nu_h}} \left(1 - \frac{E_{\nu_l}}{E_{\nu_h}}   \right)\,,
\end{equation}
such that the h.c. contributions produce light neutrinos with harder spectra than the h.f. ones.

\section{Statistical analysis} \label{sec:analysis}
We present here the Bayesian analysis we performed. Its goal is to quantify the ability of ongoing and upcoming DSNB experiments to distinguish between the no-decay case on one hand and the case where neutrino decay with a long, a medium, or a short lifetime-to-mass on the other.

For a given experiment and detection channel denoted by $j$, we consider a binned likelihood of the form
\begin{equation} \label{eq:single_likelihood}
    \mathcal{L}_{j}\left(\tau/m|\{n_i\}\right)=\prod_{i}^{N_\text{bin}}\frac{\mu_{i}^{n_{i}}}{n_{i}!}e^{-\mu_{i}} \ , 
\end{equation}
where $\mu_{i}$ represents the expected number of counts in bin $i$. The quantity $\mu_i$ depends on the given supernova simulation, decay scenario, lifetime-to-mass parameter of interest $\tau/m$ and the experimental backgrounds. The set $n_i$ denotes the \(N_\text{samp}\) binned (pseudo)data samples. Note that while \(N_\text{bin}\), \(\mu_i\) and \(n_i\) all depend on \(j\), we do not explicitly state their \(j\)-dependence for readability.

To account for the uncertainties in the signal and background, we introduce two nuisance parameters with Gaussian priors, \(\alpha\) and \(\beta\), respectively. We also combine the different detection channels and experiments by multiplying the corresponding likelihoods, to be able to make an assessment on neutrino decay when taking advantage of the cumulated statistics and different spectral distortions. For the expected signal and background in bin \(i\) given by \(s_i\) and \(b_i\), respectively, the Poisson means can then be written as \(\mu_i=(1+\alpha)s_i+(1+\beta)b_i\) and the likelihood becomes after marginalization\footnote{Note that we truncate the Gaussian distributions for \(\alpha\) and \(\beta\) at \(-1\) to avoid negative contributions. For this reason, one needs an additional normalization prefactor, which we do not explicitly write for compactness in eq.~(\ref{eq:marginalized_likelihood}).}
\begin{equation} \label{eq:marginalized_likelihood}
    \mathcal{L}_{\text{m}}\left(\tau/m\right|\{n_i\}) \propto \int_{-1}^\infty\mathrm{d}\alpha\int_{-1}^\infty\mathrm{d}\beta \left(\prod_j\prod_{i}^{N_\text{bin}}\frac{\mu_{i}^{n_{i}}}{n_{i}!}e^{-\mu_{i}}\right)\cdot\frac{e^{-\frac{\alpha^{2}}{2\sigma_{\alpha}^{2}}}}{\sqrt{2\pi}\sigma_{\alpha}}\frac{e^{-\frac{\beta^{2}}{2\sigma_{\beta}^{2}}}}{\sqrt{2\pi}\sigma_{\beta}}\,.
\end{equation}
In our analysis, we distinguish two scenarios for the uncertainties on the signal and the background: a conservative scenario with \(\sigma_\alpha=40\%\) and \(\sigma_\beta=20\%\) as in ref.~\cite{Tabrizi_2021}, as well as an optimistic scenario with \(\sigma_\alpha=20\%\) and \(\sigma_\beta=10\%\).

To quantify how well two lifetime-to-mass ratios \((\tau/m)_1\) and \((\tau/m)_0\) with equal priors \(P((\tau/m)_1)=P((\tau/m)_0)\) can be distinguished, we consider the Bayes factor
\begin{equation} \label{eq:bayes_factor}
    B_{10} = \frac{\mathcal{L}_\text{m}\left((\tau/m)_1|\{n_i\}_1\right)}{\mathcal{L}_\text{m}\left((\tau/m)_0|\{n_i\}_1\right)}\,,
\end{equation}
where \(\{n_i\}_1\) contains pseudodata generated via a Metropolis-Hastings Markov chain Monte Carlo (MCMC) algorithm, assuming lifetime-to-mass \((\tau/m)_1\). The ratio corresponds, via Bayes' theorem, to the ratio of posterior probabilities, i.e. the probabilities of hypotheses given data. Bayes factors can be interpreted with the criteria by Kass and Raftery~\cite{Kass_1995} shown in table~\ref{tab:bayes_factor_interpretation}.
\begin{table}[htbp]
    \centering
    \begin{tabular}{cc}\hline
        \(\log B_{10}\) & Level of evidence against the null hypothesis \\\hline
        \(0-1\) & Not worth more than a bare mention \\
        \(1-3\) & Positive \\
        \(3-5\) & Strong \\
        \(>5\) & Very strong \\
         \hline
    \end{tabular}
    \caption{Interpretation of logarithmic Bayes factors~\cite{Kass_1995}.}
    \label{tab:bayes_factor_interpretation}
\end{table}
As in ref.~\cite{Saez_2024} we obtain the expected evidence level to discriminate two hypotheses, i.e. here two lifetime-to-mass ratios, in future experiments by taking the mean logarithmic Bayes factor \(\langle \log B_{10}\rangle\) over samples (\(N_\text{samp}=10^6\)) generated via the MCMC algorithm\footnote{Among these samples, we apply a burn-in phase on the first \(10\%\).}. We then say that one expects (very) strong evidence against the null hypothesis when \(\langle \log B_{10}\rangle>3\) (\(\langle \log B_{10}\rangle>5\)).
two-body
\section{Numerical results on the DSNB flux and event rates} \label{sec:numerical_fluxes_and_rates}

Let us now discuss the DSNB fluxes and the corresponding event rates across different experiments. Our computations include the running SK-Gd, the upcoming JUNO, HK and DUNE experiments. As for the neutrino fluxes from core-collapse supernova simulations at a given redshift, we employ  both the Garching and Nakazato simulations.

\subsection{Predictions on the DSNB flux}
We present here DSNB predictions in presence and absence of decay.
In order to obtain the DSNB fluxes when neutrinos decay, one has to consider  a given decay scheme, choose the corresponding branching ratios, implement the energy spectra \(\psi_{ik}(E_{\nu}', E_{\nu})\)~(\ref{eq:qji}) and perform the integrals in eq.~(\ref{eq:solution})\footnote{See appendix~B in ref.~\cite{IvanezBallesteros_2023} for explicit expressions.} taking $z = 0$. The relic flux \(\phi_{\nu_i}\)  \eqref{eq:flux} is then obtained from the density \(n_{\nu_i}\) via integration over the progenitor masses (and a factor of \(c\)), with \(n_{\nu_i}^\text{NS}\) (\(n_{\nu_i}^\text{BH}\)) obtained by replacing \(Y_{\nu_i}\) by \(Y_{\nu_i}^\text{NS}\) (\(Y_{\nu_i}^\text{BH}\)) in eq.~(\ref{eq:solution}).

In our analysis we do not consider a specific model of non-radiative two-body decay to keep our results general. The majority of the results we present assume all daughter neutrinos to be active, or in other words, visible. This holds if neutrinos are Majorana particles. On the contrary, if neutrinos are Dirac particles one can have scenarios in which the daughter neutrinos can be either active or sterile (see refs.~\cite{DeGouvea_2020,IvanezBallesteros_2023} for a discussion). The case in which the daughter neutrino is invisible will be discussed at the end.
So, unless otherwise stated, the numerical results shown are for either no decay or for neutrino non-radiative decay into active neutrinos. 

The DSNB flux for flavor \(\nu_\alpha\) is given by
\begin{equation}\label{eq:projection}
    \phi_{\nu_\alpha}(E_\nu) = \sum_i\left|U_{\alpha i}\right|^2\phi_{\nu_i}(E_\nu)\,, 
\end{equation}
since the flavor and mass bases are related by $|\nu_\alpha\rangle = \sum_i U_{\alpha i}^*|\nu_i\rangle$ ($\alpha=e,\mu,\tau$ 
and $i=1,2,3$) with $U$ the Pontecorvo-Maki-Nakagawa-Sakata matrix. In our calculations, we consider the mixing angles $\theta_{23} = 49\degree, \theta_{12}=34\degree$ and $\theta_{13}=8.5\degree$ \cite{rpp_2024}.

\subsubsection{DSNB flux in absence of decay}
\begin{figure}[htbp]
    \centering
    \includegraphics[trim=0 0 0 20, clip, width=0.90\textwidth]{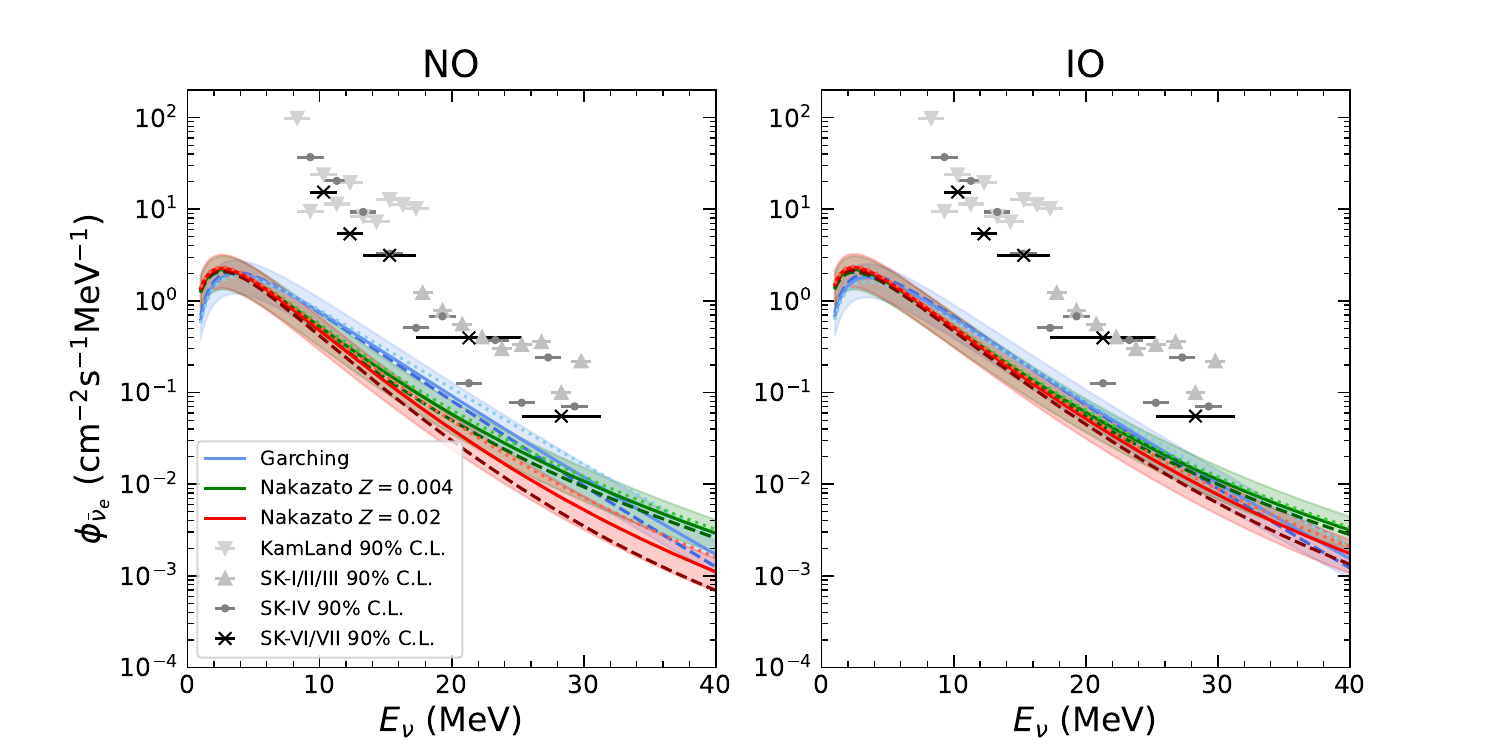}
    \caption{Relic flux of $\bar{\nu}_e$ in case of no decay for normal (left) and inverted (right) ordering. The dashed, solid and dotted lines correspond to the results obtained either for
    Garching simulations with \(f_{\text{BH}}=0.09, 0.21\text{~and~}0.41\), and to \(t_\text{rev}=100,200\text{~and~}\SI{300}{\milli\second}\) for Nakazato. The bands show the uncertainty due to the local core-collapse supernova rate relative to the solid lines. The data points correspond to the upper bounds ($90\%$ C.L.) from KamLAND~\cite{Abe_2022}, SK-I/II/III~\cite{Zhang_2015}, SK-IV~\cite{Abe_2021} and SK-VI/SK-VII \cite{Harada_2024}.}
    \label{fig:compareRelicFluxAllSimNoDecay}
\end{figure}

Let us first discuss the DSNB flux when neutrinos do not decay. The results for $\bar\nu_e$ (shown as an example) present significant variations when considering the different scenarios for the NS- and BH-contributions based on the two sets of detailed supernova simulations (figure~\ref{fig:compareRelicFluxAllSimNoDecay}).
As expected, higher BH-forming core-collapse fractions correspond to harder relic fluxes.
Moreover, for the case of inverted neutrino mass ordering the DSNB fluxes are slightly hotter than for normal ordering, because of the MSW effect (\ref{eq:MSW_IO})
and the relation \(|U_{e3}|^2\ll|U_{e2}|^2\ll|U_{e1}|^2\). Concerning the results based on Nakazato's simulations, one can see that
DSNB predictions associated with shorter shock revival times correspond to softer spectra because the accretion phase is shorter~\cite{Nakazato_2015}.

Table~\ref{tab:integrated_fluxes_no_decay} presents the integrated DSNB fluxes along with the current experimental bounds. The differences between the predictions reflect the differences across simulations seen in figure~\ref{fig:compareRelicFluxAllSimNoDecay}, due to progenitor dependence, black hole fraction and supernova (neutrino) simulations (the same evolving core-collapse supernova rate being used in the two sets). One can see that even in the small subset of detailed supernova simulations considered in this work, there is for example up to a factor of 3-4 difference,  for $\bar{\nu}_e$, between the most pessimistic Nakazato and the most optimistic Garching simulations (normal ordering).
\begin{table}[htbp]
\centering
    \begin{tabular}{|c|c||cc|c|}
        \hline
        Flavor & Ordering & Garching & Nakazato & Exp. bound \\\hline
        \(\bar\nu_e\) & NO & 0.68\,-\,0.98  & 0.26\,-\,0.63 & \multirow{2}{*}{2.7}\\
       \(E_\nu>\SI{17.3}{\mega\electronvolt}\) & IO & 0.62\,-\,0.72 & 0.40\,-\,0.63 & \\\hline
        \(\nu_e\) & NO & 0.18\,-\,0.23 & 0.12\,-\,0.21 & \multirow{2}{*}{19}\\
         \(E_\nu\in(22.9, 36.9)\)\,\si{\mega\electronvolt} & IO & 0.15\,-\,0.22 & 0.09\,-\,0.20 & \\\hline
        \(\nu_\mu+\nu_\tau\) & NO & 1.02\,-\,1.45 & 0.54\,-\,1.18 & \multirow{2}{*}{\((1.0\,$-$\,1.4)\times10^3\)}\\
         \(E_\nu>\SI{17.3}{\mega\electronvolt}\)  & IO & 1.09\,-\,1.44 & 0.62\,-\,1.20 & \\
         \hline
        \(\bar\nu_\mu+\bar\nu_\tau\) & NO & 1.27\,-\,1.56 & 0.74\,-\,1.26 & \multirow{2}{*}{\((1.3\,$-$\,1.8)\times10^3\)}\\
         \(E_\nu>\SI{17.3}{\mega\electronvolt}\) & IO & 1.33\,-\,1.81 & 0.59\,-\,1.26 & \\
         \hline
    \end{tabular}
    \caption{Experimental bounds at \(90\%\) CL and numerical results for the no-decay integrated DSNB flux (\(\si{\per\centi\metre\squared\per\second}\)) for the different flavors. The integration ranges for the neutrino energy are shown in the first column. The \(\bar\nu_e\) bound comes from a SK analysis~\cite{Abe_2021} while the \(\nu_e\) bound was obtained with SNO data~\cite{Aharmim_2020}. Additionally, analysis of SK-I data provided the bounds on \(\nu_\mu+\nu_\tau\) and \(\bar\nu_\mu+\bar\nu_\tau\)~\cite{Lunardini_2008}. Two flux values separated by a hyphen (-) represent the extrema across the different cases in the Garching or Nakazato sets.}
    \label{tab:integrated_fluxes_no_decay}
\end{table}

\begin{figure}
    \centering
    \includegraphics[trim=0 0 0 20, clip, width=0.85\linewidth]{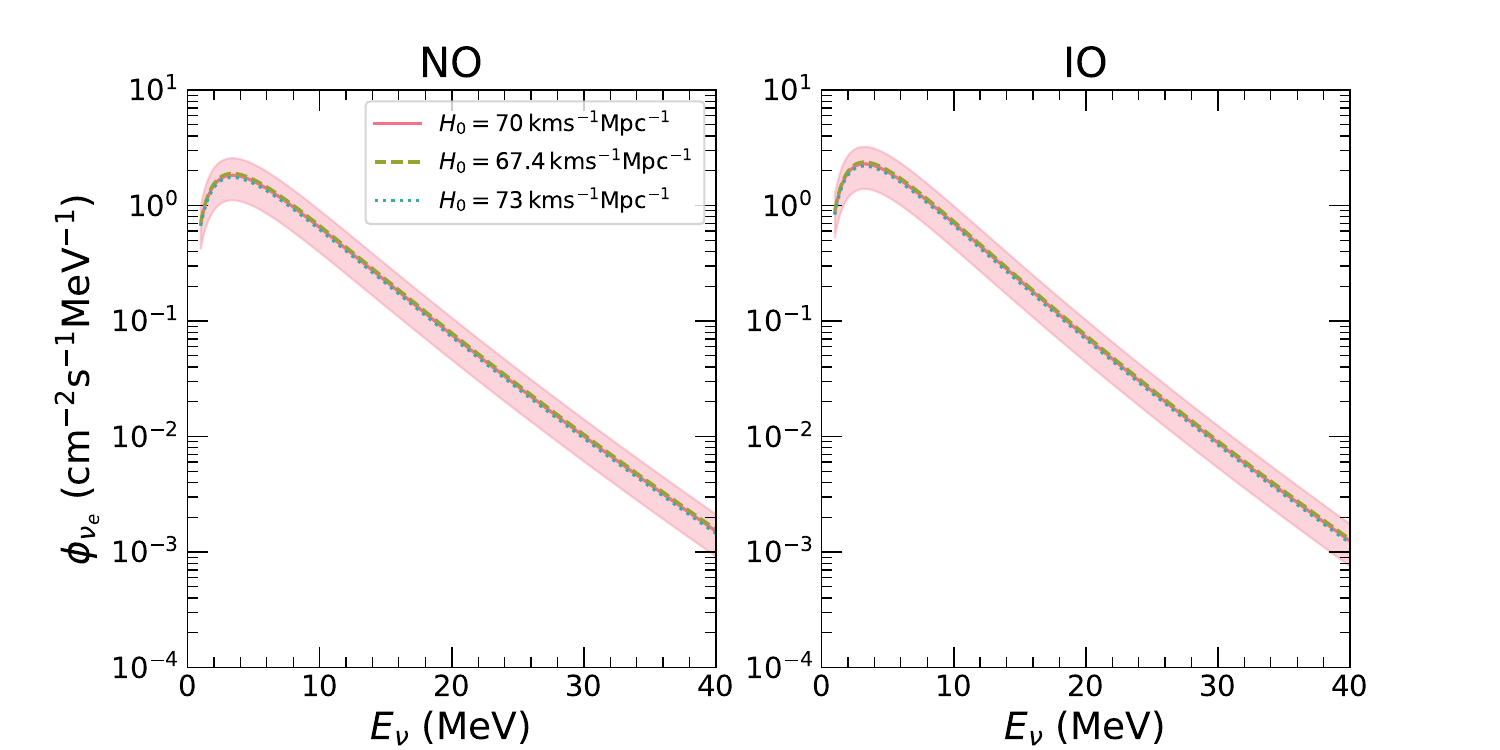}
    \caption{Relic $\nu_e$ flux in case of no decay for normal (left) and inverted (right) ordering for different values of \(H_0\). The solid, dashed and dotted lines correspond to the experimental values \(\SI{70}{\kilo\meter\per\second\per\mega\parsec}\)~\cite{Abbot_2017}, \(\SI{67.4}{\kilo\meter\per\second\per\mega\parsec}\)~\cite{Birrer_2020}, and \(\SI{73}{\kilo\meter\per\second\per\mega\parsec}\)~\cite{Riess_2022}, respectively. The results correspond to our reference simulation. The bands show the uncertainty due to the local core-collapse supernova rate relative to the solid lines.}
    \label{fig:H0_effect}
\end{figure}

Concerning the comparison with the experimental DSNB upper limits, as in ref.~\cite{IvanezBallesteros_2023}, while our integrated $\bar{\nu}_e$ fluxes are lower than the current
limit, the $\nu_e$ and $\nu_x$ ones are well below. Note that in our calculations, we also varied the Hubble constant value taking into account the current tensions on $H_0$~\cite{rpp_2024}. We found that this has a small impact on the DSNB predictions compared to the astrophysical uncertainties implemented in the present work, that come from the supernova simulations and the evolving core-collapse supernova rate (figure~\ref{fig:H0_effect}). 

\subsubsection{DSNB flux in presence of decay}
Let us now consider the possibility that neutrinos decay in vacuum when they reach the supernova surface. As in ref.~\cite{IvanezBallesteros_2023}, we take three values for  the lifetime-to-mass ratio, namely
\begin{equation}
        \left(\tau/m\right)_\text{short}=10^9\,\si{\second\per\electronvolt}
        \,\mathrm{,}\hspace{0.25cm}
        \left(\tau/m\right)_\text{medium}=10^{10}\,\si{\second\per\electronvolt}
        \hspace{0.25cm}\mathrm{and}\hspace{0.25cm}
        \left(\tau/m\right)_\text{long}=10^{11}\,\si{\second\per\electronvolt}\,.
\end{equation}
These correspond to the case of almost complete decay (``short"), to an intermediate case (``medium") and the extreme case (``long") in which neutrinos have a lab-frame lifetime close to the age of the universe $E\tau/m\sim1/H_0$ with \(E\sim O(10)\,\si{\mega\electronvolt}\)~\cite{Fogli_2004}.

\begin{figure}[htbp]
    \centering
    \includegraphics[trim=0 0 0 18, clip, width=1.0\textwidth]{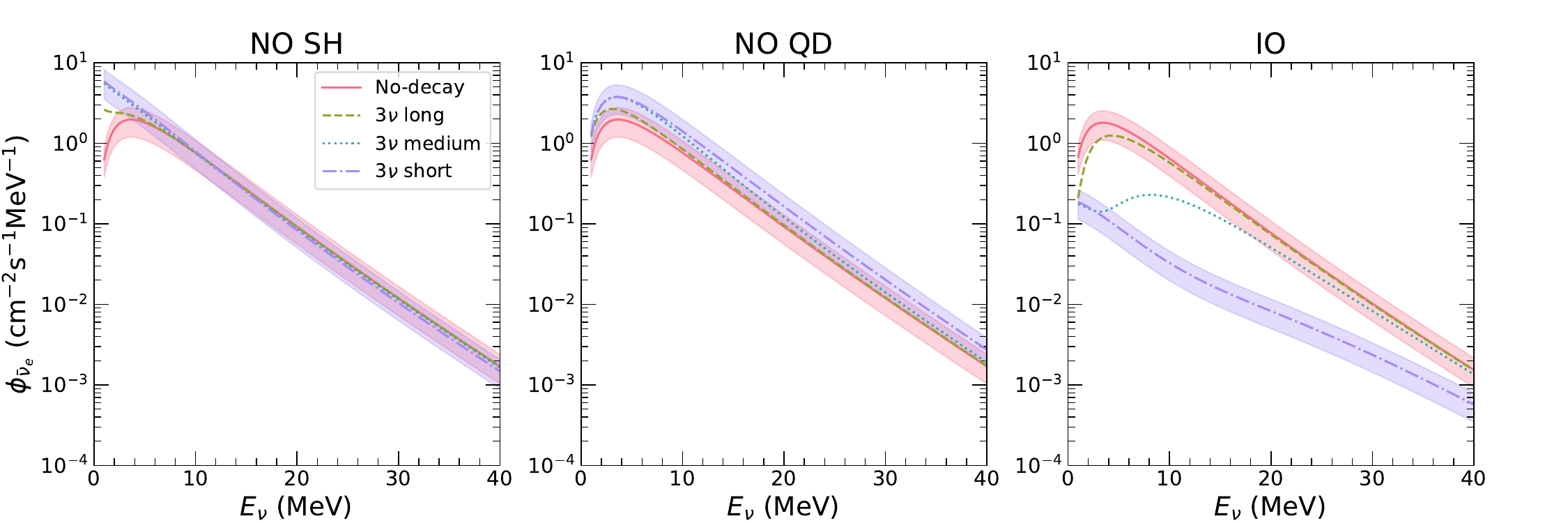}
    \caption{Relic flux of \(\bar\nu_e\) in the NO SH (left), NO QD (middle), and IO (right) patterns for the reference scenario, namely the Garching simulations with \(f_\text{BH}=0.21\). The bands around the solid and dot-dashed lines represent the uncertainty on the local core-collapse supernova rate for the no-decay and the short lifetime-to-mass cases, respectively (see text).}
    \label{fig:compareRelicFluxGarchingDecay}
\end{figure}

Our results on the fluxes\footnote{Note that as in figure~\ref{fig:compareRelicFluxAllSimNoDecay} we only show here results for $\bar{\nu}_e$ since the results for $\nu_e$ show similar behaviours.} are shown in figure~\ref{fig:compareRelicFluxGarchingDecay} and are the same as those in ref.~\cite{IvanezBallesteros_2023} in a full 3$\nu$ flavor framework when we use the same DSNB inputs, in particular the Garching simulations, the same scenarios for the progenitor dependence and parametrization (and uncertainties) for the evolving core-collapse supernova rate. For consistency and comparison with the findings of ref.~\cite{IvanezBallesteros_2023} we take the same reference scenario, namely the Garching simulations with \(f_\text{BH}=0.21\). Unless stated otherwise, results are shown assuming this reference scenario. Note however that we performed calculations for the other black hole fractions and for the Nakazato simulations as well, and found analogous qualitative behaviors.

In agreement with the findings of ref.~\cite{IvanezBallesteros_2023}, the dependence of the DSNB fluxes on \(\tau/m\) varies greatly across the three mass patterns. In the NO SH case, the fluxes are degenerate among each other and with the no-decay predictions (except for $E_{\nu} < \SI{5}{\mega\electronvolt}$). In the NO QD case the DSNB fluxes with decay are enhanced compared to the ones without (cfr. figure~7 of ref.~\cite{IvanezBallesteros_2023} where the reference scenario shown is for $f_{\rm BH} = 0.21$), while for IO the inclusion of neutrino non-radiative decay suppresses significantly the flux for $(\tau/m)_{\rm short}$ and also for $(\tau/m)_{\rm medium}$ for $E_{\nu} < \SI{20}{\mega\electronvolt} $. 
The flux suppression for IO is explained by\footnote{An analogous argument holds for the \(\nu_e\) flux (not shown).} the decay of the \(\bar\nu_1\) mass eigenstate into the \(\nu_3\) and \(\bar\nu_3\) and the smallness of \(|U_{e3}|^2\). On the other hand, the enhancement for NO QD is due to the decay of the heavier mass states into the \(\bar\nu_1\) state (see figure~\ref{fig:decay_patterns}).

The integrated fluxes when neutrino decay is included are given in tables~\ref{tab:integrated_fluxes_decay_garching} and~\ref{tab:integrated_fluxes_decay_nakazato} for the Garching and the Nakazato simulations respectively. Obviously similar trends are found for each of the mass patterns.

\begin{table}[htbp]
\centering
    \begin{tabular}{|c|c||c|ccc|}
        \hline
        Experiment & Pattern & No decay & \((\tau/m)_\text{long}\) & \((\tau/m)_\text{medium}\) & \((\tau/m)_\text{short}\)\\\hline
         SK-Gd, JUNO & NO SH & 1.79\,-\,2.28 & 1.77\,-\,2.27 & 1.69\,-\,2.20  & 1.64\,-\,2.17\\
        (\(\bar\nu_e\)) & NO QD & 1.79\,-\,2.28 & 1.89\,-\,2.39 & 2.53\,-\,3.06 & 3.34\,-\,3.89\\
        \((12.8, 30.8)\)\,\si{\mega\electronvolt} & IO & 1.62\,-\,1.68 & 1.52\,-\,1.58 & 0.93\,-\,0.96 & 0.14\,-\,0.15\\\hline
        HK & NO SH & 0.65\,-\,0.91 & 0.64\,-\,0.90 & 0.61\,-\,0.87 & 0.57\,-\,0.84\\
        (\(\bar\nu_e\)) & NO QD & 0.65\,-\,0.91 & 0.67\,-\,0.94 & 0.85\,-\,1.15 & 1.18\,-\,1.52\\
        \((17.3, 31.3)\)\,\si{\mega\electronvolt} & IO & 0.59\,-\,0.67 & 0.56\,-\,0.64 & 0.40\,-\,0.45 & 0.07\,-\,0.08\\\hline
        DUNE & NO SH & 0.39\,-\,0.46 & 1.59\,-\,2.05 & 1.51\,-\,1.99 & 1.46\,-\,1.96\\
        (\(\nu_e\)) & NO QD & 0.39\,-\,0.46 & 0.40\,-\,0.48 & 0.48\,-\,0.60 & 0.66\,-\,0.87\\
        \((19, 31)\)\,\si{\mega\electronvolt} & IO & 0.35\,-\,0.46 & 0.34\,-\,0.44 & 0.24\,-\,0.32 & 0.05\,-\,0.06\\
         \hline
    \end{tabular}
    \caption{Integrated DSNB flux (\(\si{\per\centi\metre\squared\per\second}\)) in absence and presence of neutrino decay for different values of the lifetime-to-mass ratios. The energy ranges used in the integrals are shown underneath the acronym of each experiment. In each case, the two flux values separated by a hyphen (-) represent the extrema across the different cases based on the Garching simulations.}
    \label{tab:integrated_fluxes_decay_garching}
\end{table}

\begin{table}[htbp]
\centering
    \begin{tabular}{|c|c||c|ccc|}
        \hline
        Experiment & Pattern & No decay & \((\tau/m)_\text{long}\) & \((\tau/m)_\text{medium}\) & \((\tau/m)_\text{short}\)\\\hline
        SK-Gd, JUNO & NO SH & 0.74\,-\,1.38 & 0.73\,-\,1.37 & 0.68\,-\,1.32  & 0.64\,-\,1.30\\
        (\(\bar\nu_e\)) & NO QD & 0.74\,-\,1.38 & 0.80\,-\,1.46 & 1.20\,-\,2.01 & 1.72\,-\,2.69\\
        \((12.8, 30.8)\)\,\si{\mega\electronvolt} & IO & 1.01\,-\,1.35 & 0.96\,-\,1.28 & 0.58\,-\,0.77 & 0.08\,-\,0.11\\\hline
        HK & NO SH & 0.24\,-\,0.55 & 0.23\,-\,0.54 & 0.22\,-\,0.52 & 0.19\,-\,0.51\\
        (\(\bar\nu_e\)) & NO QD & 0.24\,-\,0.55 & 0.25\,-\,0.57 & 0.36\,-\,0.75 & 0.57\,-\,1.06\\
        \((17.3, 31.3)\)\,\si{\mega\electronvolt} & IO & 0.36\,-\,0.54 & 0.35\,-\,0.52 & 0.25\,-\,0.36 & 0.04\,-\,0.06\\\hline
        DUNE & NO SH & 0.24\,-\,0.38 & 0.24\,-\,0.38 & 0.23\,-\,0.36 & 0.21\,-\,0.35\\
        (\(\nu_e\)) & NO QD & 0.24\,-\,0.38 & 0.25\,-\,0.40 & 0.29\,-\,0.50 & 0.37\,-\,0.71\\
         \((19, 31)\)\,\si{\mega\electronvolt}  & IO & 0.20\,-\,0.37 & 0.19\,-\,0.35 & 0.13\,-\,0.25 & 0.03\,-\,0.04\\
         \hline
    \end{tabular}
    \caption{Integrated DSNB flux (\(\si{\per\centi\metre\squared\per\second}\)) in absence and presence of neutrino decay for different values of the lifetime-to-mass ratios. The energy ranges used in the integrals are shown underneath the acronym of each experiment. In each case, the two flux values separated by a hyphen (-) represent the extrema across the different cases based on the Nakazato simulations.}
    \label{tab:integrated_fluxes_decay_nakazato}
\end{table}

\subsection{Expected DSNB rates}
Before presenting our predictions for the DSNB rates, let us summarize the experimental parameters and detection channels on which our analysis is based.

\subsubsection{Detection channels and parameters}
The experiments that have the potential to observe the DSNB are SK-Gd, HK, JUNO and DUNE.
The relevant experimental parameters for the computations of the DSNB rates, including the efficiencies and the running times, are given in table~\ref{tab:detector_parameters}.

SK is an underground water Cherenkov detector in Kamioka (Japan) with a total mass (fiducial volume) of \(\SI{50}{\kilo\tonne}\) (\(\SI{22.5}{\kilo\tonne}\))~\cite{Abe_2021}. In 2020 the SK-Gd experiment started taking data, with gadolinium (Gd) addition to improve neutron tagging. SK-Gd ran two-years with $0.01\%$ Gd concentration (SK-VI) and 
is currently running with \(0.03\%\) concentration (SK-VII). The main detection channel for the DSNB in SK is inverse beta-decay (IBD)
\begin{equation}
    \bar{\nu}_e  + p \rightarrow n + e^+\,,
\end{equation}
where the positron energy is related to the neutrino energy via \(E_{e^+} = E_{\bar\nu_e} - \Delta_{np}\), \(\Delta_{np}=\SI{1.293}{\mega\electronvolt}\) being the neutron-proton mass difference.

HK will be the largest water Cherenkov detector ever built, with \(\SI{258}{\kilo\tonne}\) (fiducial volume \(\SI{187}{\kilo\tonne}\))~\cite{Abe_2018}. Located about \(\SI{8}{\kilo\metre}\) south of SK at the Tochibora site, HK should start taking data in 2027. The inclusion of gadolinium is under study.  Thanks to the fiducial volume increase by a factor of 8.3 with respect to SK, other channels might become relevant for DSNB measurements in HK, such as neutrino-electron elastic scattering 
\begin{equation}
    \nu+e^-\rightarrow\nu+e^-
\end{equation}
or CC interactions with oxygen
\begin{equation}
    \bar\nu_e + \ce{^16O}\rightarrow e^++\ce{^16N}^*\hspace{0.7cm}\mathrm{and}\hspace{0.7cm}\nu_e + \ce{^16O}\rightarrow e^-+\ce{^16F}^*\,,
\end{equation}
with energy thresholds of \(\SI{16}{\mega\electronvolt}\) and \(\SI{14}{\mega\electronvolt}\) respectively.
For $\nu$-$e$, the visible energy is the electron recoil kinetic energy \(E_\text{R}\), while for $\bar{\nu}_e\,$-$^{16}{\rm O}$ (${\nu}_e\,$-$\ce{^16O}$), it is the positron (electron) energy \(E_{e^+}\) (\(E_{e^-}\)), along with the energy from \(\ce{^16N}^*\) or \(\ce{^16F}^*\) decay products.

JUNO will be a liquid scintillator detector of 20 kt (18 kt fiducial volume), located in Jiangmen (South China)~\cite{An_2016}. The detector is divided into an inner (FV1, 14.7 kt) and an outer (FV2, 3.6 kt) fiducial volumes with different backgrounds and signal efficiencies~\cite{Abusleme_2022}. While the main detection channel for DSNB measurements is IBD, JUNO will also be sensitive to neutrino-proton elastic scattering first discussed in ref. \cite{Beacom_2002hs}
\begin{equation}
    \nu + p \rightarrow \nu + p\,,
\end{equation}
with visible energy being the quenched proton kinetic energy \(T'\). 

The liquid argon detector DUNE is located at the Sanford Underground Research Facility (South Dakota). Its far detector will be composed of four LAr TPC modules with 40 kt fiducial volume~\cite{Acciarri_2016}. The detector will start operating in a staged approach, from 2029 to 2034. The main detection channel is  
\begin{equation}
    {\nu}_e  + \ce{^40Ar} \rightarrow e^- +  \ce{^40K}^*\,,
\end{equation}
where the visible energy is the electron energy $E_{e^-}$ and the one of $^{40}{\rm K}^*$ decay products.

\begin{table}[htbp]
\centering
    \begin{tabular}{|c|c|cccc|c|}
        \hline
        Experiment & Channel & $N_{\rm t}$ $(10^{33})$ & $\epsilon$ $(\%)$ & $t$ (\si{\year}) & $E$ (\(\si{\mega\electronvolt}\)) & Refs.\\\hline
        SK-VI & IBD & 1.5 & 40 & 2 & (11.5, 29.5) & \cite{IvanezBallesteros_2023, Santos_2024}\\
        SK-VII & IBD & 1.5 & 55 & 8 & (11.5, 29.5) & \cite{IvanezBallesteros_2023, Santos_2024}\\
        HK & IBD & 12.5 & 30 & 20 & (16.0, 30.0) & \cite{IvanezBallesteros_2023, Santos_2024} \\
        HK-Gd & IBD & 12.5 & 40 & 20 & (16.0, 30.0) & \cite{IvanezBallesteros_2023, Santos_2024} \\
        HK, HK-Gd & $\nu$-$e$ & 62.5 & 100 & 20 & (10.0, 20.0) & \cite{Tabrizi_2021} \\
        HK, HK-Gd & $\nu_e\,$-$\ce{^16O}$ & 6.25 & 100 & 20 & (16.0, 32.0) & \cite{Kolbe_2002} \\
        HK, HK-Gd & $\bar\nu_e\,$-$\ce{^16O}$ & 6.25 & 100 & 20 & (14.0, 34.0) & \cite{Kolbe_2002} \\
        JUNO (FV1) & IBD & 0.99 & 84 & 20 & (11.5, 29.5) & \cite{Abusleme_2022}\\
        JUNO (FV2) & IBD & 0.24 & 77 & 20 & (11.5, 29.5) & \cite{Abusleme_2022}\\
        JUNO (FV1) &  $\nu$-$p$  & 0.99 & 100 & 20 & (0.2, 1.5) & \cite{Dasgupta_2011}\\
        JUNO (FV2) &  $\nu$-$p$  & 0.24 & 100 & 20 & (0.2, 1.5) & \cite{Dasgupta_2011}\\
        DUNE & $\nu_e\,$-$\ce{^40Ar}$ & 0.602 & 86 & 20 & (19, 31) & \cite{Moeller_2018, Abi_2021}\\\hline
    \end{tabular}
    \caption{Parameters characterizing the experiments and detection channels. The quantities \(N_\text{t}\), \(\epsilon\) and \(t\) refer to the number of targets, the detector efficiency and the period of data taking respectively. The column labeled by \(E\) shows the considered energy range in terms of positron energy \(E_{e^+}\) for IBD, electron recoil energy \(E_\text{R}\) for $\nu$-$e$, neutrino energy 
 \(E_\nu\) for $\nu_e\,$-$\ce{^16O}$, $\bar{\nu}_e\,$-$\ce{^16O}$ and for $\nu_e\,$-$\ce{^40Ar}$, and quenched proton kinetic energy \(T'\) for  $\nu$-$p$.}
    \label{tab:detector_parameters}
\end{table}

\subsubsection{DSNB event rates} \label{sec:rates}
\noindent
The DSNB event rate for a neutrino of flavor \(\alpha\) can be calculated from the relic flux via~\cite{Saez_2024}
\begin{equation}\label{eq:event_rates}
    \frac{\mathrm{d}N_{\nu_\alpha}}{\mathrm{d}t} = \epsilon N_\text{t} \int_{E_a}^{E_b} \mathrm{d}E'\int_{E_\text{thr}(E')}^\infty \mathrm{d}E_\nu \phi_{\nu_{\alpha}}(E_{\nu}) \frac{\mathrm{d}\sigma}{\mathrm{d}E'}(E_{\nu}, E') \,,
\end{equation}
where \(\epsilon\) is the efficiency, \(N_\text{t}\) the number of targets, 
\(E'\) is the measured energy and \(\frac{\mathrm{d}\sigma}{\mathrm{d}E'}(E_{\nu}, E')\) the differential cross-section in a given detection channel. The neutrino energy threshold is \(E_\text{thr}(E')\), while \(E_a\), \(E_b\) are the bounds of the \(E'\) energy bin. 

For the IBD channel, where \(E'=E_{e^+}\), the integral over \(E_\nu\) is immediate. We use \(\sigma^\text{IBD}(E_\nu)\) obtained via the \(E_\nu<\SI{300}{\mega\electronvolt}\) expression from ref.~\cite{Strumia_2003}.
For neutrino-electron elastic scattering, we use the cross-section \(\frac{\mathrm{d\sigma}}{\mathrm{d}E_\text{R}}(E_\nu, E_\mathrm{R})\) from ref.~\cite{DeGouvea_2020_neutrino_electron}.
For the \(\nu_e\,\)-\(\ce{^16O}\), \(\bar\nu_e\,\)-\(\ce{^16O}\) and \(\nu_e\,\)-\(\ce{^40Ar}\) channels, we replace the differential cross section by the total cross-sections as a function of neutrino energy from ref.~\cite{snowglobes_webpage}.

Last, the \(\nu\)-\(p\) channel requires a specific treatment because the protons being slow, the scintillation light they produce in the detector is quenched. This feature can be described via the quenching function
\begin{equation}
    T'(T) = \int_0^T \frac{\mathrm{d}T}{1 + k_B \left\langle \frac{\mathrm{d}T}{\mathrm{d}x} \right\rangle}\,,
\end{equation}
where \(k_B\) is the Birks constant~\cite{Birks_1951}, \(T\) the (unquenched) proton kinetic energy and \(T'\) the quenched proton kinetic energy. We used the numerical values of the quenching function from ref.~\cite{vonKrosigk_2013}.
For the differential cross-section \(\frac{\mathrm{d}\sigma}{\mathrm{d}T}(T,E_\nu)\), we use in our calculations the expression to zeroth order in \(E_\nu/m_p\) from ref.~\cite{Dasgupta_2011}. Because of quenching, eq.~(\ref{eq:event_rates}) needs to be adjusted to
\begin{equation}
    \frac{\mathrm{d}N_{\nu_\alpha}}{\mathrm{d}t} = \epsilon N_\text{t} \int_{T'_a}^{T'_b}\mathrm{d}T'\frac{\mathrm{d}T}{\mathrm{d}T'}\int_{E_\text{thr}(T)}^{\infty}\mathrm{d}E_{\nu} \phi_{\nu_{\alpha}}(E_{\nu}) \frac{\mathrm{d}\sigma}{\mathrm{d}T}(E_\nu,T)\,,
\end{equation}
where \(E_\text{thr}(T)=\sqrt{m_pT/2}\)~\cite{vonKrosigk_2013}.

\begin{figure}[htbp]
    \centering
    \includegraphics[trim=0 0 0 20, clip, width=0.75\textwidth]{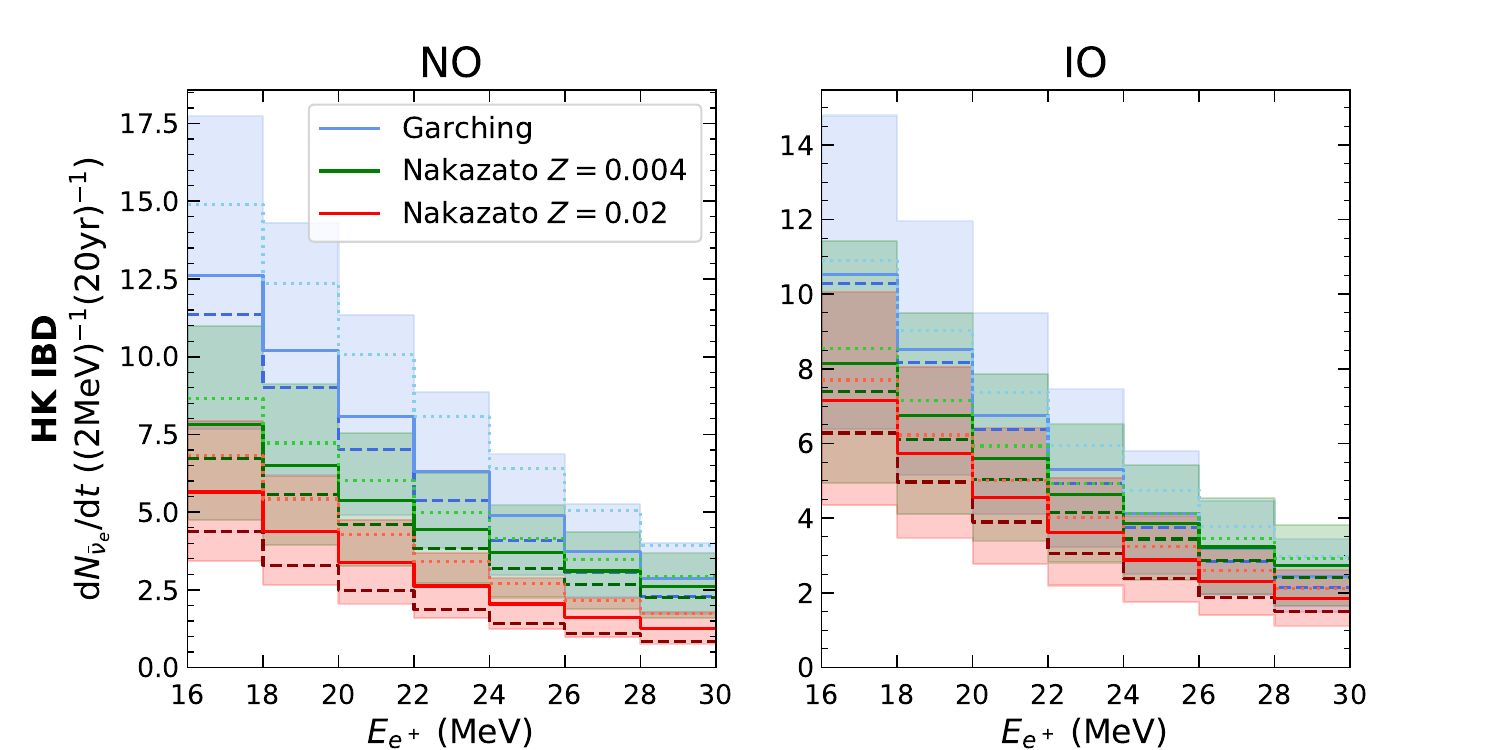}
    \caption{Expected event rates for the IBD channel in HK in the cases of normal (left) and inverted (right) neutrino mass ordering for the different simulations. The results shown are in absence of neutrino decay. The dashed, solid and dotted lines correspond respectively to \(f_{\text{BH}}=0.09, 0.21\text{~and~}0.41\) for Garching, and to \(t_\text{rev}=100,200\text{~and~}\SI{300}{\milli\second}\) for Nakazato. The bands show the uncertainty due to the local core-collapse supernova rate relative to the results given by the solid lines.}
    \label{fig:HK_IBD_no_decay_rates_all_sim}
\end{figure}

Predictions on the DSNB event rates are given in absence of decay for the four experiments, using both Nakazato and Garching simulations using the scenarios shown in figures~\ref{fig:nakazato_templates} and~\ref{fig:garching_templates} respectively. The integrated number of events are shown in table~\ref{tab:integrated_rates_no_decay} for all detectors and channels. 
\begin{table}[htbp]
\centering
    \begin{tabular}{|c|c|c|cc|}
        \hline
        Experiment & Channel & Ordering & Garching & Nakazato \\\hline
        \multirow{2}{*}{SK-Gd} & \multirow{2}{*}{IBD} & NO & 10\,-\,13  & 4\,-\,8\\
         & & IO & 9 & 5\,-\,8\\\hline
        \multirow{2}{*}{HK} & \multirow{2}{*}{IBD} & NO & 51\,-\,73 & 18\,-\,45\\
         & & IO & 46\,-\,54 & 29\,-\,45\\\hline
        \multirow{2}{*}{HK-Gd} & \multirow{2}{*}{IBD} & NO & 68\,-\,98 & 25\,-\,60\\
         & & IO & 62\,-\,72 & 38\,-\,59\\\hline
        \multirow{2}{*}{HK, HK-Gd} & \multirow{2}{*}{$\nu$-$e$} & NO & 3 & 1\,-\,2\\
         & & IO & 3 & 1\,-\,2\\\hline
        \multirow{2}{*}{HK, HK-Gd} & \multirow{2}{*}{$\nu_e\,$-$\ce{^16O}$} & NO & 1 & 0\,-\,1\\
         & & IO & 0\,-\,1 & 0\,-\,1\\\hline
        \multirow{2}{*}{HK, HK-Gd} & \multirow{2}{*}{$\bar{\nu}_e\,$-$\ce{^16O}$} & NO & 1 & 0\,-\,1\\
         & & IO & 0\,-\,1 & 0\,-\,1\\\hline
        \multirow{2}{*}{JUNO} & \multirow{2}{*}{IBD} & NO & 25\,-\,33 & 10\,-\,20\\
         & & IO & 23\,-\,25 & 14\,-\,20\\\hline
        \multirow{2}{*}{JUNO} & \multirow{2}{*}{$\nu$-$p$} & NO & 4\,-\,7 & 3\,-\,8\\
         & & IO & 4\,-\,7 & 3\,-\,8\\\hline
        \multirow{2}{*}{DUNE} & \multirow{2}{*}{$\nu_e\,$-$\ce{^40Ar}$} & NO & 10\,-\,13 & 7\,-\,11\\
          & & IO & 9\,-\,12 & 5\,-\,10\\
         \hline
    \end{tabular}
    \caption{Number of DSNB events expected for the Garching and Nakazato simulations in absence of neutrino decay, using the parameters shown in table~\ref{tab:detector_parameters}. Two flux values separated by a hyphen (-) represent the extrema across the different cases in the Garching or Nakazato sets. If the results are the same, a single value is shown.}
    \label{tab:integrated_rates_no_decay}
\end{table}
One can see, as for the case of the (integrated) DSNB fluxes, that there are significant variations in the predictions, across the different scenarios considered.
This is also visible in figure~\ref{fig:HK_IBD_no_decay_rates_all_sim}, which shows the IBD event rates using different simulations for the HK experiment.
Note that the $\nu$-$e$, $\nu_e\,$-$\ce{^16O}$ and $\bar{\nu}_e\,$-$\ce{^16O}$ channels in HK and HK-Gd only yield very low event rates.

\begin{figure}[htbp]
    \centering
    \includegraphics[trim=0 0 0 10, clip, width=0.99\textwidth]{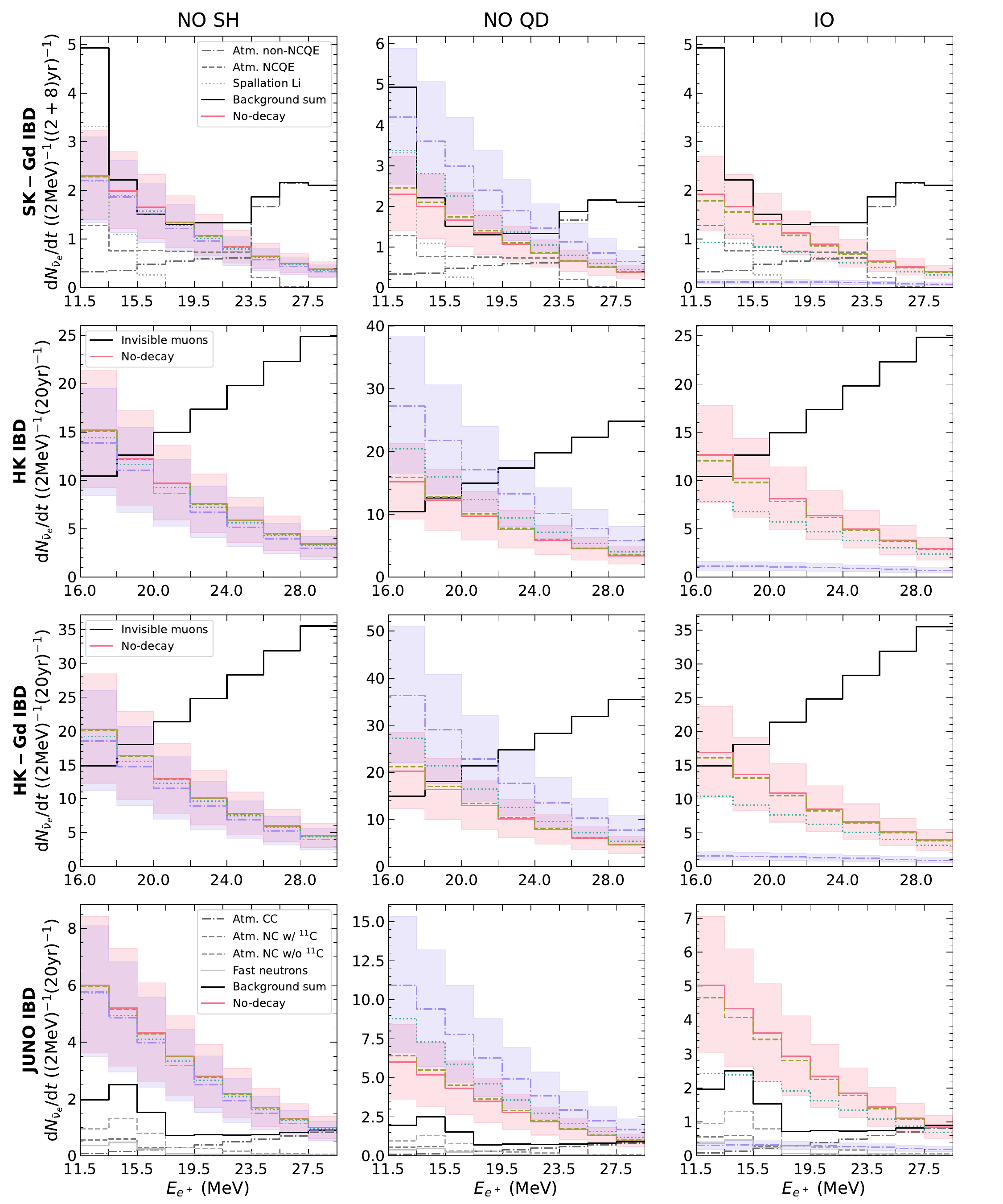}
    \caption{Number of IBD events expected in SK-Gd, HK, HK-Gd and JUNO when neutrinos undergo non-radiative two-body decay with short (dot-dashed), medium (dotted) and long (dashed lines) lifetime-to-mass ratios. The results correspond to the mass patterns NO SH (left), NO QD (middle), and IO (right) and are obtained for our reference scenario, i.e. the Garching simulations with \(f_\text{BH}=0.21\). The bands show the uncertainty on the evolving core-collapse supernova rate relative to the no-decay case and decay case with short lifetime-to-mass ratio. The background rates are also shown.}
    \label{fig:IBD_decay_event_rates}
\end{figure}

Let us now consider the DSNB predictions when neutrinos decay non-radiatively. The total expected number of events, integrated over the energy, are given in table~\ref{tab:integrated_rates_decay} for the long and short lifetime-to-mass ratios; whereas figure~\ref{fig:IBD_decay_event_rates} shows the IBD event rates  in SK-Gd, HK and HK-Gd and JUNO, along with the associated backgrounds. 

First of all, as one can see from figure~\ref{fig:IBD_decay_event_rates} a similar trend in the predicted number of events is visible, as expected, in the four experiments.
Such trends differ for the two neutrino mass ordering and for the different mass patterns, in agreement with ref.~\cite{IvanezBallesteros_2023}. 
Obviously they reflect the DSNB flux behaviours seen in figure~\ref{fig:compareRelicFluxGarchingDecay}.
For NO SH, the predicted events in absence of decay are degenerate with those with decay and lifetime-to-mass ratios in the range \(\tau/m \in [10^9, 10^{11}]\,\si{\second\per\electronvolt}\). In the NO QD
case,  the rates with decay and $(\tau/m)_{\rm short}$ are increased with respect to $(\tau/m)_{\rm medium}$ and $(\tau/m)_{\rm long}$, the latter being essentially degenerate with the no-decay case. On the contrary, for IO, the DSNB events for $(\tau/m)_{\rm short}$ are suppressed by up to a factor of 10 at low energies, when compared to predictions in absence of neutrino decay. Such a suppression is milder but still significant even for $(\tau/m)_{\rm medium}$.
  
Since the backgrounds in the different experiments will be an element of our Bayesian investigation we  also discuss them here briefly for each of the experiment considered.
The most important source of background for IBD in SK-Gd are neutral-current quasi-elastic (NCQE) and non-NCQE atmospheric backgrounds. The latter includes charged-current (CC) contributions from atmospheric \(\nu_e\) and \(\bar\nu_e\) as well as invisible muons (that fall below the Cherenkov threshold and whose decay electrons produce a background). As reported in ref.~\cite{Abe_2021}, atmospheric non-NCQE sources dominate. An additional background at low energy comes from decaying $^9$Li isotope produced by spallation of oxygen nuclei induced by cosmic ray muons. Following~\cite{Santos_2024}, since the non-NCQE background as well as other backgrounds have been significantly reduced, for our computations we take SK-VI/VII backgrounds by scaling those from ref.~\cite{Santos_2024_poster} to 2 years and ref.~\cite{Harada_2024} to 8 years.

In the case of HK, the background is dominated by invisible muons, for which we used the numerical values of figure~188 (right)\footnote{Unlike sometimes assumed in the literature, the backgrounds presented in this figure are not expected to hold for HK-Gd, but for HK with neutron capture on protons~\cite{Santos_2024}).} in ref.~\cite{Abe_2018} and scaled them for expected neutron tagging efficiencies of \(35\%\) and \(50\%\) in HK and HK-Gd, respectively~\cite{Santos_2024}. As for the other background sources that include atmospheric NC and atmospheric \(\bar\nu_e\) CC we take them to be negligible compared to invisible muons in the region of interest (due to the progress performed in the SK-Gd experiment, and assuming this can be translated to the operational HK), see~\cite{Abe_2021, Harada_2023, Santos_2024_poster, Harada_2024}.

As for JUNO, pulse shape discrimination techniques allow for a substantially enhanced signal-over-noise ratio, see ref.~\cite{Abusleme_2022}. We take the IBD background estimates from this recent analysis. In particular, after pulse-shape discrimination, atmospheric CC interactions remain the main source of background. Additional ones include fast neutrons associated with untagged muons and NC atmospheric neutrino interactions on \(^{11}\text{C}\) and other targets. Note that \(^{11}\text{C}\) contributions can be distinguished thanks to the associated beta-decay signature.  Concerning neutrino-proton scattering events in JUNO, as in ref.~\cite{Dasgupta_2011}, backgrounds (e.g. from radioactivity) are expected to be very high in the range \(T'\in(0.2, 1.5)\,\si{\mega\electronvolt}\) over which we compute the event rates.
Finally, for the DUNE detector, we follow ref.~\cite{Moeller_2018} and assume backgrounds from atmospheric CC interactions similar to those in the analogous LAr ICARUS detector~\cite{Cocco_2004} since DSNB backgrounds in DUNE are still under investigation.

\begin{table}[htbp]
\centering
    \begin{tabular}{|c|c|c|cc|cc|}
        \hline
        \multirow{2}{*}{Experiment} & \multirow{2}{*}{Channel} & \multirow{2}{*}{Scenario} & \multicolumn{2}{c|}{Garching} & \multicolumn{2}{c|}{Nakazato} \\
         & & & $(\tau/m)_{\rm long}$ & $(\tau/m)_{\rm short}$ & $(\tau/m)_{\rm long}$ & $(\tau/m)_{\rm short}$\\\hline
        \multirow{3}{*}{SK-Gd} & \multirow{3}{*}{IBD} & NO SH & 10\,-\,13 & 9\,-\,12 & 4\,-\,8 & 3\,-\,7\\
         & & NO QD & 10\,-\,13 & 18\,-\,22 & 4\,-\,8 & 9\,-\,15 \\
         & & IO & 8\,-\,9 & 1 & 5\,-\,7 & 1 \\\hline
        \multirow{3}{*}{HK} & \multirow{3}{*}{IBD} & NO SH & 50\,-\,73 & 45\,-\,67 & 18\,-\,45 & 15\,-\,42\\
         & & NO QD & 53\,-\,75 &  93\,-\,122 & 20\,-\,47 & 45\,-\,86 \\
         & & IO & 45\,-\,52 & 6\,-\,7 & 28\,-\,43 & 4\,-\,5\\\hline
        \multirow{3}{*}{HK-Gd} & \multirow{3}{*}{IBD} & NO SH & 67\,-\,97 & 59\,-\,89 & 24\,-\,60 & 20\,-\,55\\
         & & NO QD & 70\,-\,101 & 123\,-\,163 & 26\,-\,63 & 60\,-\,115 \\
         & & IO & 59\,-\,69 & 9\,-\,10 & 37\,-\,57 & 5\,-\,7\\\hline
        \multirow{3}{*}{JUNO} & \multirow{3}{*}{IBD} & NO SH & 25\,-\,33 & 23\,-\,31 & 10\,-\,20 & 8\,-\,19\\
         & & NO QD & 26\,-\,35 & 46\,-\,57 & 11\,-\,22 & 23\,-\,39\\
         & & IO & 22\,-\,23 & 2\,-\,3 & 14\,-\,19 & 1\,-\,2\\\hline
        \multirow{3}{*}{DUNE} & \multirow{3}{*}{$\nu_e\,$-$\ce{^40Ar}$} & NO SH & 10\,-\,12 & 9\,-\,11 & 7\,-\,11 & 6\,-\,10\\
         & & NO QD & 11\,-\,13 & 17\,-\,23 & 7\,-\,11 & 10\,-\,20\\
         & & IO & 9\,-\,12 & 1\,-\,2 & 5\,-\,10 & 1\\
        \hline
    \end{tabular}
    \caption{Expected number of DSNB events in the main detection channel of each experiment. The results are obtained including neutrino two-body non-radiative decay
    for different decay mass patterns with either long or short lifetime-to-mass ratios, using the parameters shown in table~\ref{tab:detector_parameters}. The two flux values separated by a hyphen (-) represent the extrema across the different cases in the Garching or Nakazato sets of simulations and scenarios. If the results are the same, a single value is shown.}
    \label{tab:integrated_rates_decay}
\end{table}

\begin{figure}[htbp]
    \centering
    \includegraphics[scale=0.47]{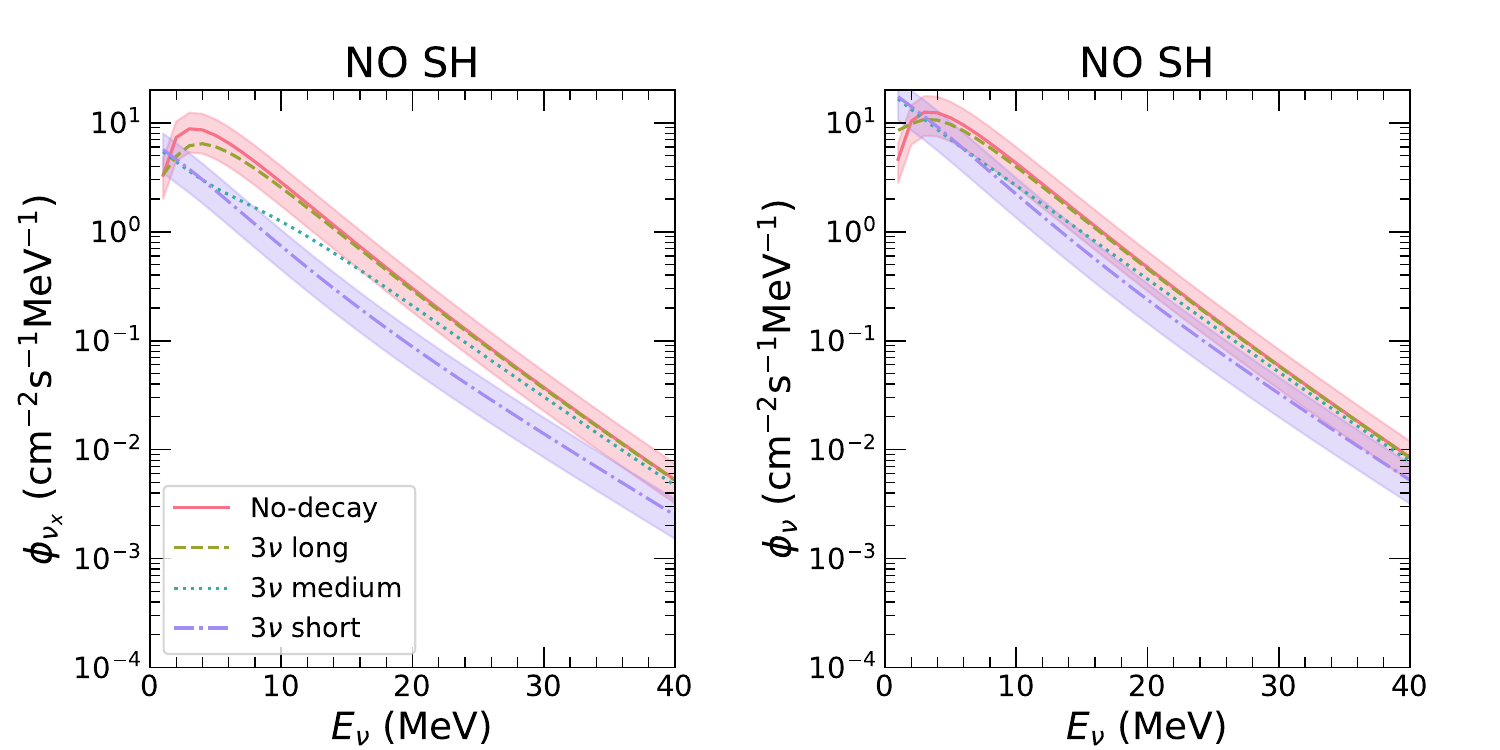}
    \caption{DSNB expected fluxes for $\nu_x$ (left) and for the all neutrino flavors (right) in presence of neutrino decay. The bands show the uncertainty on the evolving core-collapse supernova rate relative to the no-decay case and to the decay case with a short lifetime-to-mass ratio.}
    \label{fig:nup-nux}
\end{figure}

\begin{figure}[htbp]
    \centering
    \includegraphics[trim=0 10 0 10, clip, width=0.99\textwidth]{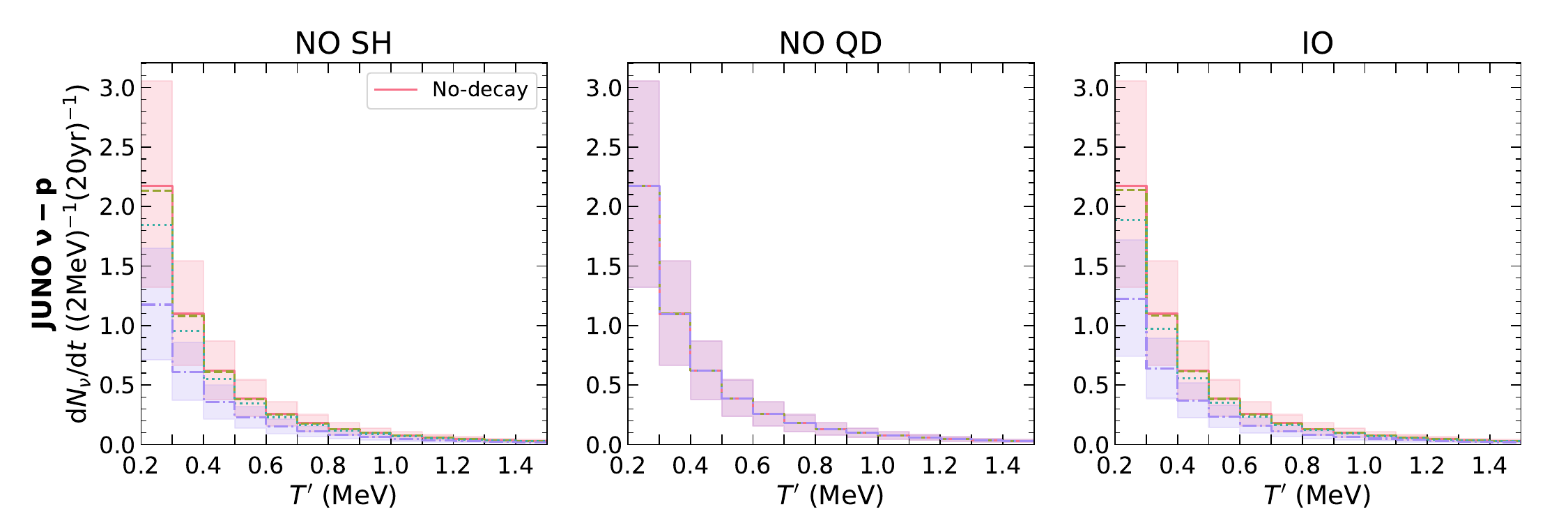}
    \caption{DSNB expected events in $\nu$-$p$ channel in JUNO detector when neutrino decay is included with short (dot-dashed), medium (dotted) and long (dashed lines) lifetime-to-mass ratios. The results are for the decay mass patterns NO SH (left), NO QD (middle), and IO (right), and correspond to our reference scenario, i.e. the Garching simulations with \(f_\text{BH}=0.21\). The bands show the uncertainty on the evolving core-collapse supernova rate relative to the no-decay case and decay case with short lifetime-to-mass ratio.}
    \label{fig:JUNO_nu-p_decay_event_rates}
\end{figure}

JUNO could also use the  $\nu$-$p$  channel~\cite{Dasgupta_2011} for the DSNB, as discussed in refs.~\cite{Tabrizi_2021}. Our predictions for the expected event rates
are shown in figure~\ref{fig:JUNO_nu-p_decay_event_rates}. The absolute number of events is low for our set of simulations: we expect \(5\) events if neutrinos do not decay, and \(3\) for $(\tau/m)_{\rm short}$ in the Garching simulation with \(f_\text{BH}=0.21\) for the NO SH mass pattern.  Since the channel is sensitive to the sum over all neutrino flavors, the only possible source of degeneracy breaking in the event rates comes here from energy loss during decay, see eq.~(\ref{eq:solution}). 
Interestingly, one observes that the event rates in the NO QD scenario are completely degenerate in the  $\nu$-$p$  channel, due to the Dirac-shaped energy spectrum, eq.~(\ref{eq:qd_spectrum}); whereas the degeneracies found in the IBD channel in the NO SH scenario are not present. The latter is due to the fact that neutrino decay has a different impact on the electron  and the non-electron neutrino flavors which contribute with a different weight to the total neutrino flux (figure \ref{fig:nup-nux}). 
These results suggest that neutrino-proton scattering could offer a potential path to break the degeneracies between DSNB predictions in absence and in presence of neutrino decay in the NO SH case. To this aim, a much larger number of $\nu$-proton events would be necessary while reducing expected backgrounds at a very low level.

\begin{figure}[htbp]
    \centering
    \includegraphics[trim=0 10 0 10, clip, width=0.99\textwidth]{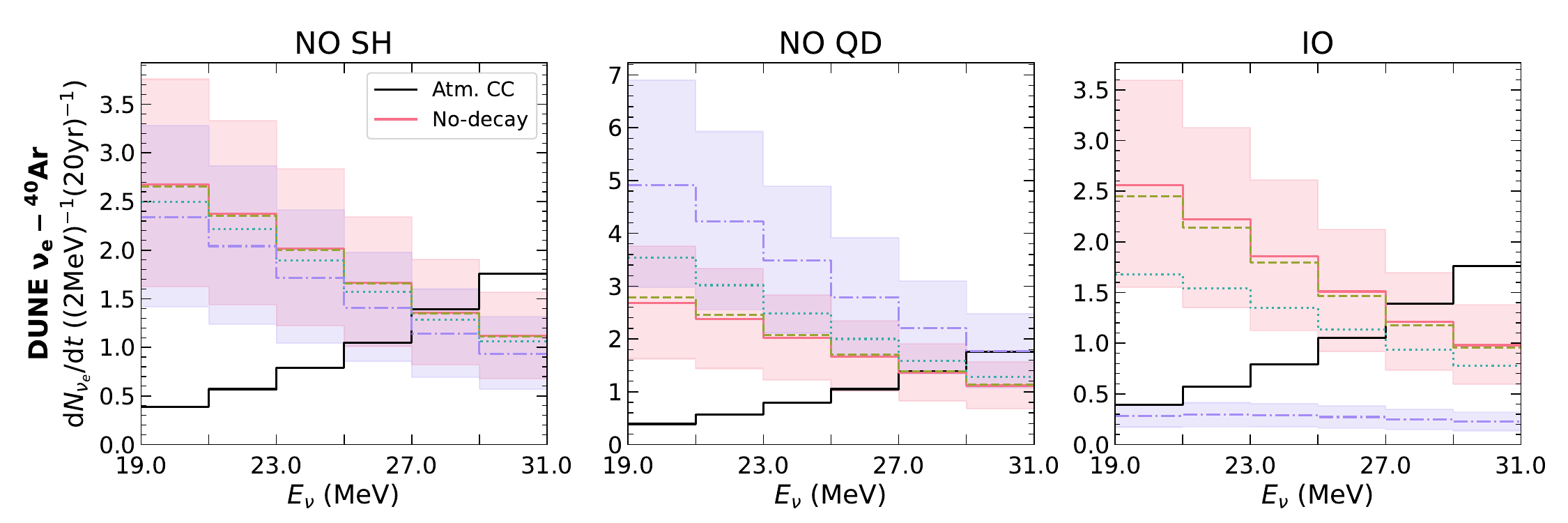}
    \caption{DSNB expected events in the $\nu$-$\ce{^40Ar}$ detection channel in DUNE when neutrino decay is included with short (dot-dashed), medium (dotted) and long (dashed lines) lifetime-to-mass ratios. The results are for the decay mass patterns NO SH (left), NO QD (middle), and IO (right), and correspond to our reference scenario, i.e. the Garching simulations with \(f_\text{BH}=0.21\). The bands show the uncertainty on the evolving core-collapse supernova rate relative to the no-decay case and decay case with short lifetime-to-mass ratio. Background rates are also shown following ref. \cite{Moeller_2018}.}
    \label{fig:DUNE_Ar_decay_event_rates}
\end{figure}

Last, figure~\ref{fig:DUNE_Ar_decay_event_rates} shows the expected event rates associated with $\nu_e\,$-$\ce{^40Ar}$ scattering in DUNE. 
Backgrounds are also shown following ref. \cite{Moeller_2018}.
Although the signal-to-noise ratio seems promising, right now little information is available on what the background in DUNE will actually be by the time of data taking. 
 
\section{Breaking the degeneracies?} \label{sec:analysis_results}
We now turn to the Bayesian analysis to investigate our ability to discriminate between the case in which neutrino do not decay non-radiatively and the one in which they do. To this aim we first discuss whether each experiment, taken individually, would be sensitive enough and consider different lifetime-to-mass ratios. We then combine the detection channels and the experiments in order to explore how much improvement can be obtained. Unless stated otherwise, we present results assuming the reference scenario, i.e. Garching simulations with \(f_\text{BH}=0.21\). We will also employ a conservative (optimistic) scenario that assumes uncertainties of \(40\%\) (\(20\%\)) on the signal and \(20\%\) (\(10\%\)) on the background.

\subsection{Considering each experiment separately}
\begin{figure}[htbp]
    \centering
    \begingroup
    \captionsetup[subfigure]{skip=-0.6em}
    \begin{subfigure}{0.49\textwidth}
        \centering
        \includegraphics[trim=0 30 0 90, clip, width=\textwidth]{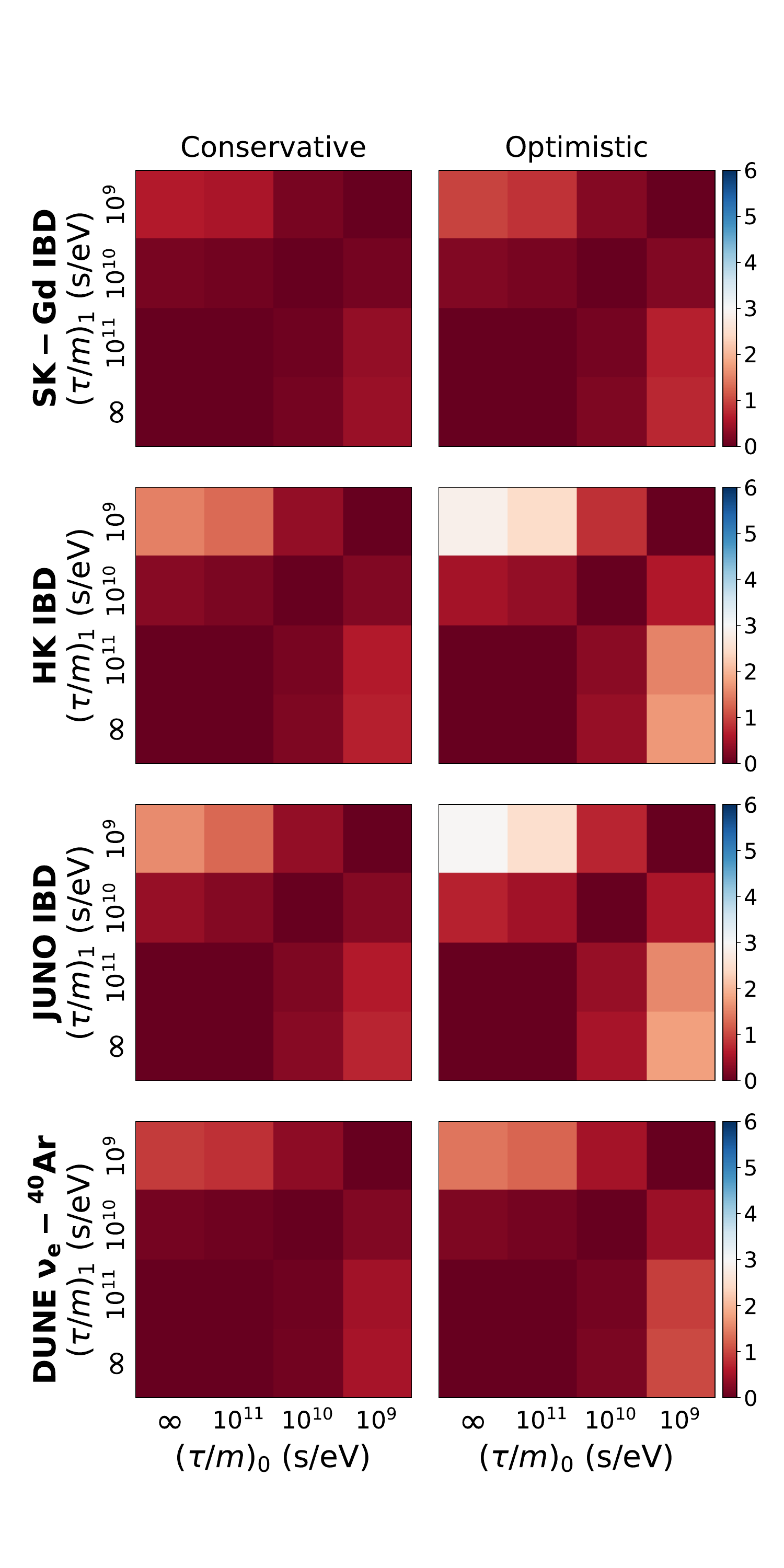}
        \caption{NO QD}
        \label{fig:bayes_factors_main_channels_bg_conservative_optimistic_fBH021_NO_QD}
    \end{subfigure}
    \hfill
    \begin{subfigure}{0.49\textwidth}
        \centering
        \includegraphics[trim=0 30 0 90, clip, width=\textwidth]{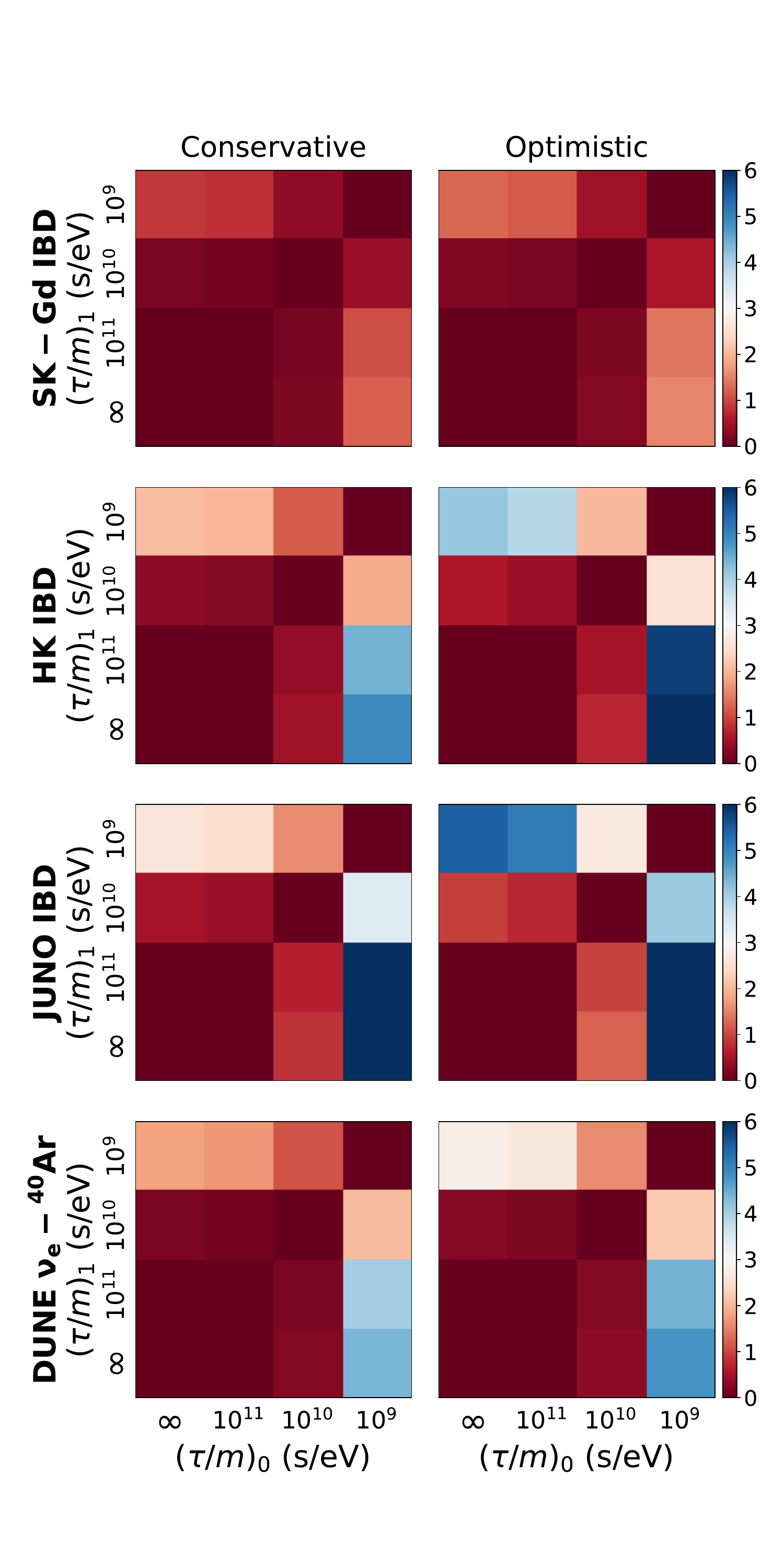}
        \caption{IO}
        \label{fig:bayes_factors_main_channels_bg_conservative_optimistic_fBH021_IO}
    \end{subfigure}
    \endgroup
    \caption{The mean logarithmic Bayes factors \(\langle \log B_{10}\rangle\) are shown for each experiment considering the IBD for SK-Gd, HK and JUNO or $\nu_e\,$-$\ce{^40Ar}$ scattering in DUNE for the (a) NO QD or the (b) IO mass patterns. Results are shown for the conservative (left, 40 $\%$ on the signal, \(20\%\) on the background) or optimistic (right, \(20\%\) on the signal, \(10\%\) on the background) scenarios on the uncertainties.}
    \label{fig:bayes_factors_main_channels_bg_conservative_optimistic_fBH021}
\end{figure}

Our ability to disentangle the DSNB predictions without decay from those that include neutrino decay,  in a given experiment, strongly depends on the neutrino decay mass pattern.
Indeed, in the NO SH scenario, the degeneracies found in the DSNB fluxes and rates are such that the logarithmic Bayes factors almost vanish in all experiments considered in the IBD and $^{40}$Ar detection channels. Obviously the situation is better in the NO QD and IO scenarios as one could expect from the behaviour of the DSNB fluxes. Figure~\ref{fig:bayes_factors_main_channels_bg_conservative_optimistic_fBH021} shows the mean logarithmic Bayes factor \(\langle \log B_{10}\rangle\) and presents interesting results for our conservative and optimistic scenarios for the uncertainties in the signal and backgrounds in each of the four experiments. 
\indent

For the currently running SK-Gd experiment, as expected from the behavior of the event rates (figure~\ref{fig:IBD_decay_event_rates}), no pair of lifetime-to-mass ratios can be distinguished with strong evidence in the NO QD mass pattern, while the case \(({\tau/m})_{\rm short}=10^9\,\si{\second\per\electronvolt}\) appears the easiest to distinguish. This is even more so for IO and with optimistic uncertainties, because of the \(\bar\nu_e\) flux suppression. Such trends are general and appear in the IBD results for all experiments. 

Thanks to the larger IBD statistics gained in the upcoming HK experiment, strong evidence against stable neutrinos can almost be reached if $({\tau/m})_1=10^9 $\(\,\si{\second\per\electronvolt}\) for the NO QD pattern when assuming optimistic uncertainties. On the other hand, for IO the hypothesis ${(\tau/m})_0 \lesssim 10^{9} $\(\,\si{\second\per\electronvolt}\) could be rejected with (very) strong evidence if neutrinos are stable, or decay with \(({\tau/m})_1=10^{11}\,\si{\second\per\electronvolt}\) in the conservative (optimistic) scenario. Additionally, in the optimistic scenario, strong evidence against neutrinos decaying with $({\tau/m})_0 \gtrsim 10^{11}$\(\,\si{\second\per\electronvolt}\) is expected  if $(\tau/m)_1 \lesssim 10^9$\(\,\si{\second\per\electronvolt}\). Contrarily, the low number of events in the $\nu$-$e$, $\nu_e\,$-$\ce{^16O}$ and $\bar{\nu}_e\,$-$\ce{^16O}$ channels yields vanishingly small Bayes factors and therefore does not influence the combined analysis (as we verified).

As for JUNO, thanks to pulse-shape discrimination of the background,  the performance is similar to the one of HK for the NO QD mass pattern, while it is definitely better for IO.
Indeed, for the latter, the results show that very strong evidence for (against) stable neutrinos against (for) $({\tau/m})_0 \lesssim 10^{9} $\(\,\si{\second\per\electronvolt}\)
($(\tau/m)_1 \lesssim 10^{9}$\(\,\si{\second\per\electronvolt}\)) can be reached in the conservative (optimistic) case.  Finally, for the DUNE experiment we find that while strong evidence for or against even 
$(\tau/m)_{\rm short}$ cannot be reached for normal ordering, strong evidence against it could be obtained in inverted ordering if neutrinos do not decay, or decay with $({\tau/m})_1=10^{11} $\(\,\si{\second\per\electronvolt}\) (figure~\ref{fig:bayes_factors_main_channels_bg_conservative_optimistic_fBH021}). 

\begin{figure}
    \centering
    \includegraphics[width=0.99\textwidth]{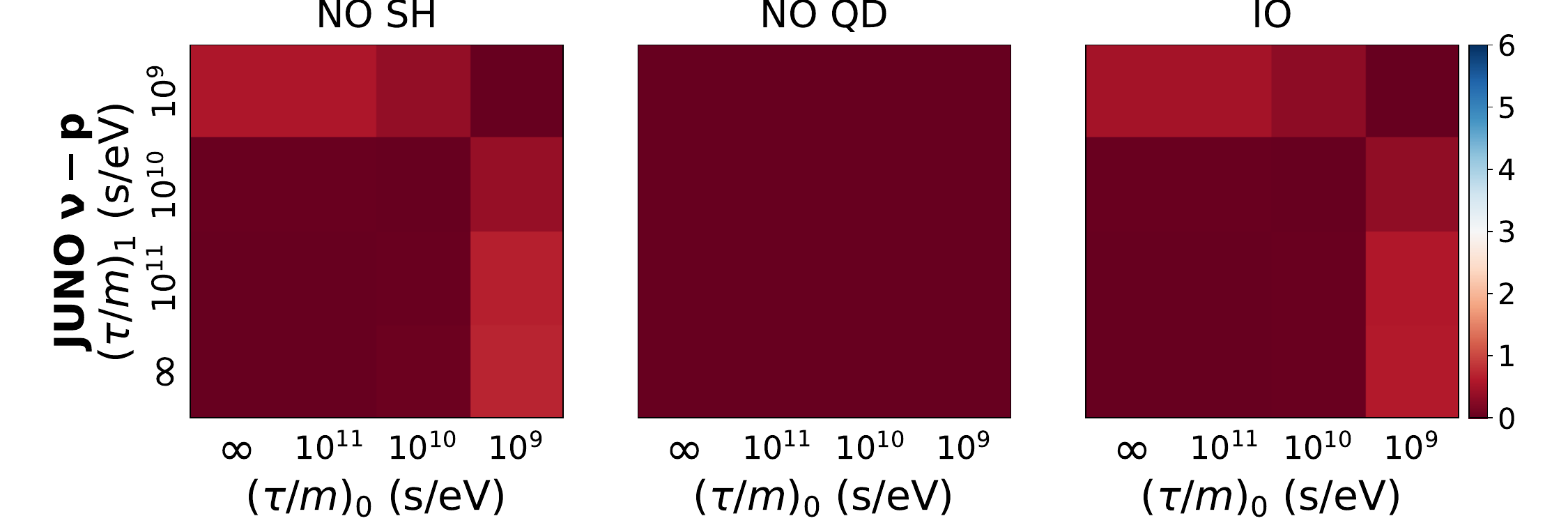}
    \caption{Mean logarithmic Bayes factors \(\langle\log B_{10}\rangle\) for NO SH (left), NO QD (centre) and IO (right) scenarios with the JUNO \(\nu\)-\(p\) channel. Both backgrounds and uncertainties are neglected. Note that backgrounds (e.g. from radioactivity) in this channel are expected to be very large.}
\label{fig:bayes_factors_JUNO_nu_p_bg_no_fBH021}
\end{figure}

Additionally we investigated as possible detection channel $\nu$-$p$ scattering, which was discussed in ref.~\cite{Tabrizi_2021} in relation with the DSNB. 
We provide the Bayes factors in  figure~\ref{fig:bayes_factors_JUNO_nu_p_bg_no_fBH021} for the three mass patterns, which shows low Bayes factors for NO SH and IO on the one hand, and complete degeneracy for NO QD (as in figure~\ref{fig:JUNO_nu-p_decay_event_rates}) on the other. So, our results for NO SH are in contrast with those of  ref.~\cite{Tabrizi_2021}: the number of events being very low such detection channel does not bring any improvement.

\begin{figure}[htbp]
    \centering
    \begingroup
    \begin{subfigure}{0.7\textwidth}
        \centering
        \includegraphics[scale=0.35]{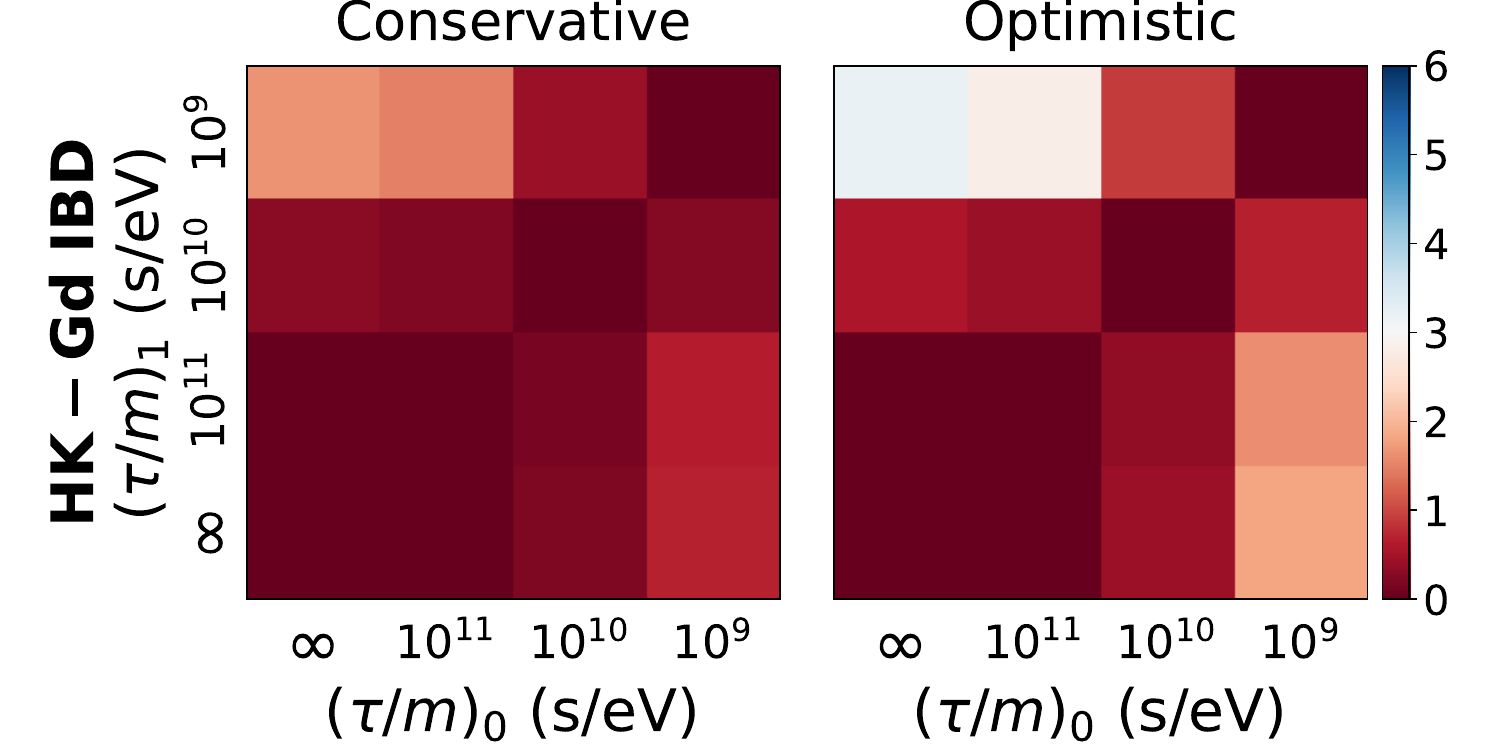}
        \caption{NO QD}
        \label{fig:bayes_factors_HK_Gd_IBD_bg_conservative_optimistic_fBH021_NO_QD}
    \end{subfigure}
    \hfill
    \begin{subfigure}{0.7\textwidth}
        \centering
        \includegraphics[scale=0.35]{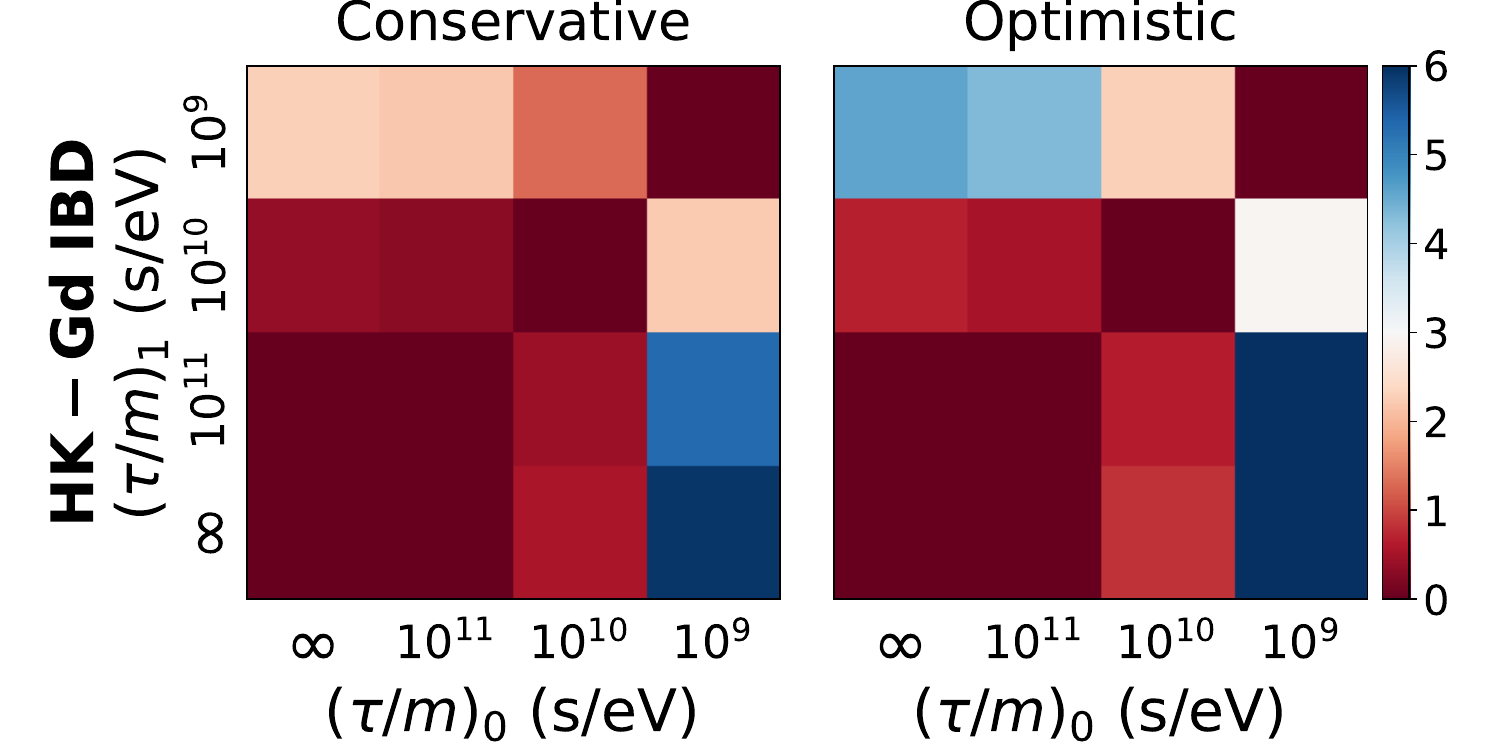}
        \caption{IO}
        \label{fig:bayes_factors_HK_Gd_IBD_bg_conservative_optimistic_fBH021_IO}
    \end{subfigure}
    \endgroup
    \caption{Mean logarithmic Bayes factor \(\langle \log B_{10}\rangle\) in the IBD channel in HK-Gd for (a) NO QD and (b) IO. Both the conservative  (left, 40 $\%$ on the signal, \(20\%\) on the background) and the optimistic (right, \(20\%\) on the signal, \(10\%\)  scenarios) for the uncertainties are considered.}
    \label{fig:bayes_factors_HK_Gd_bg_conservative_optimistic_fBH021}
\end{figure}

If HK is doped with Gd, the improvement in the neutron tagging efficiency enhances the expected Bayes factors as visible in figure~\ref{fig:bayes_factors_HK_Gd_bg_conservative_optimistic_fBH021}. 
Indeed the threshold for strong evidence against the no-decay case can be reached for $({\tau/m})_1=10^9 $\(\,\si{\second\per\electronvolt}\) for NO QD with optimistic uncertainties. For IO, evidence against $({\tau/m})_0 \lesssim 10^{9} $\(\,\si{\second\per\electronvolt}\) for stable neutrinos or $({\tau/m})_{\rm long}=10^{11} $\(\,\si{\second\per\electronvolt}\) becomes very strong even in the conservative scenario.

\begin{figure}
    \centering
    \includegraphics[trim=0 0 0 0, clip, width=0.9\linewidth]{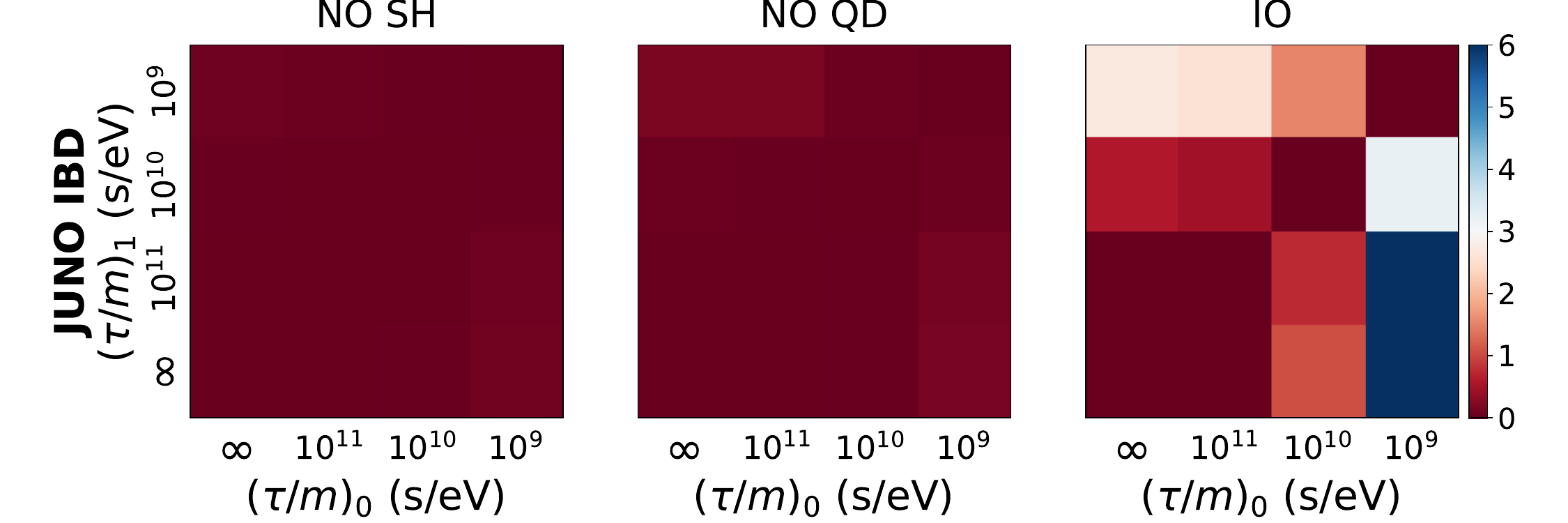}
    \caption{Mean logarithmic Bayes factors \(\langle \log B_{10}\rangle\) for IBD in JUNO with the NO SH (left), NO QD (middle) and IO (right) mass patterns, considering neutrino invisible decay. We assume democratic branching ratios between helicity flipping and helicity conserving decays (see text). Results are shown for the conservative scenario on the uncertainties (40 $\%$ on the signal, \(20\%\) on the background).}
    \label{fig:JUNO_invisible}
\end{figure}

Before presenting the combined analyses, we would like to discuss how our results change when one considers neutrino decay to invisible decay products. In order to address this issue we performed calculations considering the same decay mass patterns of figure~\ref{fig:decay_patterns}, but also considering decay into sterile neutrinos. The branching ratios assumed are again democratic, with equal branching ratios assigned to helicity flipping and helicity conserving decays.
The corresponding results of the Bayesian analysis are shown in figure~\ref{fig:JUNO_invisible} for the JUNO experiment with IBD. For NO SH, the Bayes factors remain very low. Interestingly, the NO QD case worsens when considering invisible decay. This is likely due to the visible neutrino flux loss mitigating the decay-induced enhancement. For inverted ordering, the results appear unchanged.

\subsection{Combined analysis} \label{sec:combined_analysis}
\begin{figure}[htbp]
    \centering
        \includegraphics[scale = 0.3]{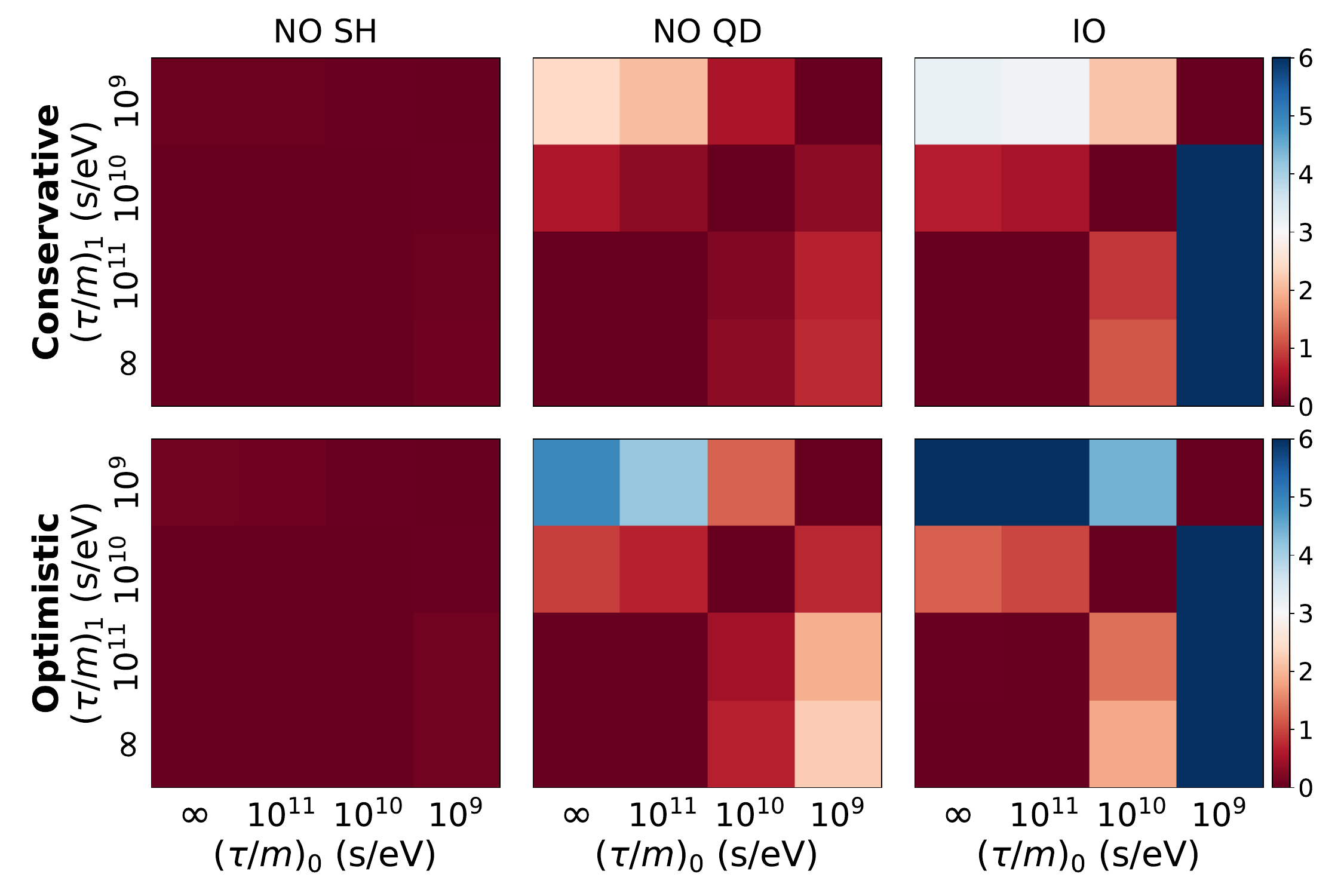}
        \label{fig:bayes_factors_combined_bg_conservative_optimistic_fBH021_HK}
    \caption{Mean logarithmic Bayes factor \(\langle\log B_{10}\rangle\) for the combined analysis of the IBD channels in SK-Gd, JUNO and HK and the $\nu_e\,$-$\ce{^40Ar}$  channel in DUNE. Results are shown for the conservative (top, \(40\%\) on the signal, \(20\%\) on the background) and optimistic (bottom, \(20\%\) on the signal, \(10\%\) on the background) scenarios for the uncertainties.}
    \label{fig:bayes_factors_combined_bg_conservative_optimistic_fBH021}
\end{figure}

Combining the experiments yields improved sensitivities on the lifetime-to-mass ratios, as compared to experiments taken separately, with one exception, namely the NO SH pattern. This can be seen in figure~\ref{fig:bayes_factors_combined_bg_conservative_optimistic_fBH021}, which presents the mean logarithmic Bayes factors for the NO SH, NO QD and IO mass patterns with the conservative and optimistic uncertainties. The combination includes the SK-Gd, JUNO, HK and DUNE experiments (note that replacing HK by HK-Gd in the combination yields the same outcomes). Combining the detection channels yields now strong evidence for $({\tau/m})_{\rm short}$ against no-decay in the optimistic scenario for NO QD.
For IO any scenario can be clearly distinguished against $(\tau/m)_0 = 10^9$\(\,\si{\second\per\electronvolt}\) for conservative and optimistic uncertainties.

\begin{figure}[htbp]
    \centering
    \includegraphics[scale=0.35]{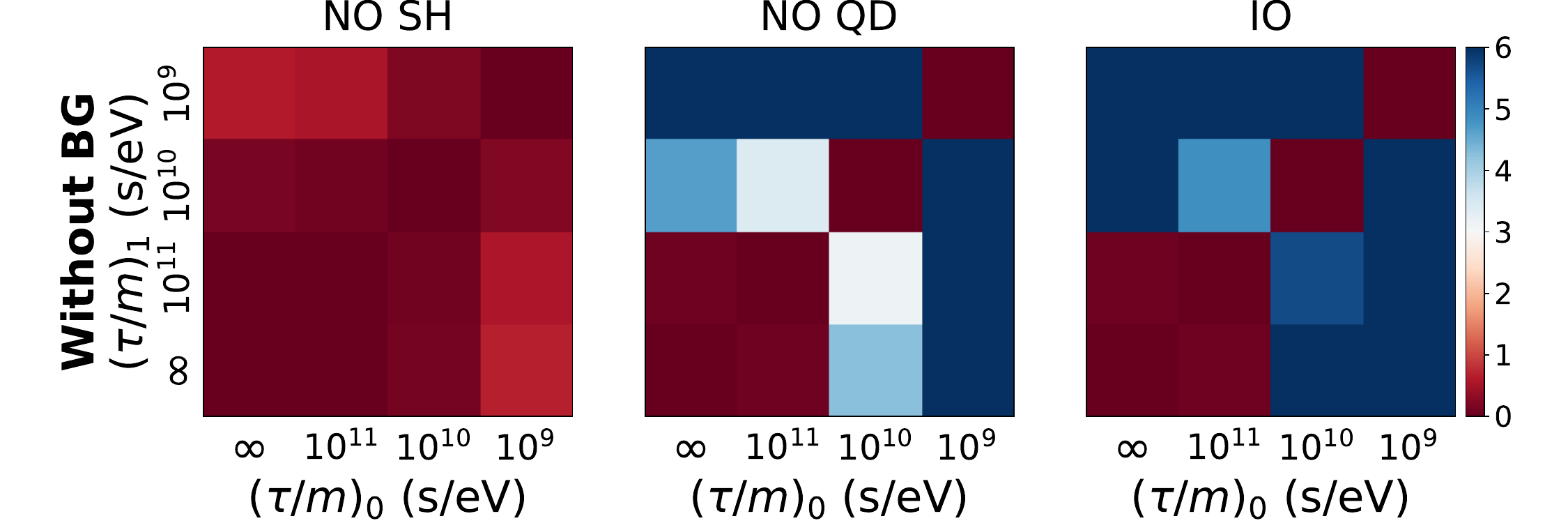}
    \caption{Mean of the logarithmic Bayes factors \(\langle\log B_{10}\rangle\) for the combined analysis of the IBD channels in SK-Gd, JUNO and HK and the $\nu_e\,$-$\ce{^40Ar}$ channel in DUNE in a completely idealized case that assumes vanishing uncertainties on the DSNB flux and neglecting backgrounds.}
    \label{fig:bayes_factors_combined_no_uncertainties_fBH021_HK}
\end{figure}

Figure~\ref{fig:bayes_factors_combined_no_uncertainties_fBH021_HK} shows what could be the Bayes factors at the best, when there is an idealized situation when negligible uncertainties in the flux and no backgrounds. One can see that even in this idealized situation, the Bayes factors are practically the same for NO SH in absence and presence of decay. Also, the case of no decay cannot be distinguished from the long lifetime-to-mass ratio whatever mass pattern is considered.

\begin{figure}[htbp]
    \centering
     \includegraphics[scale=0.35]{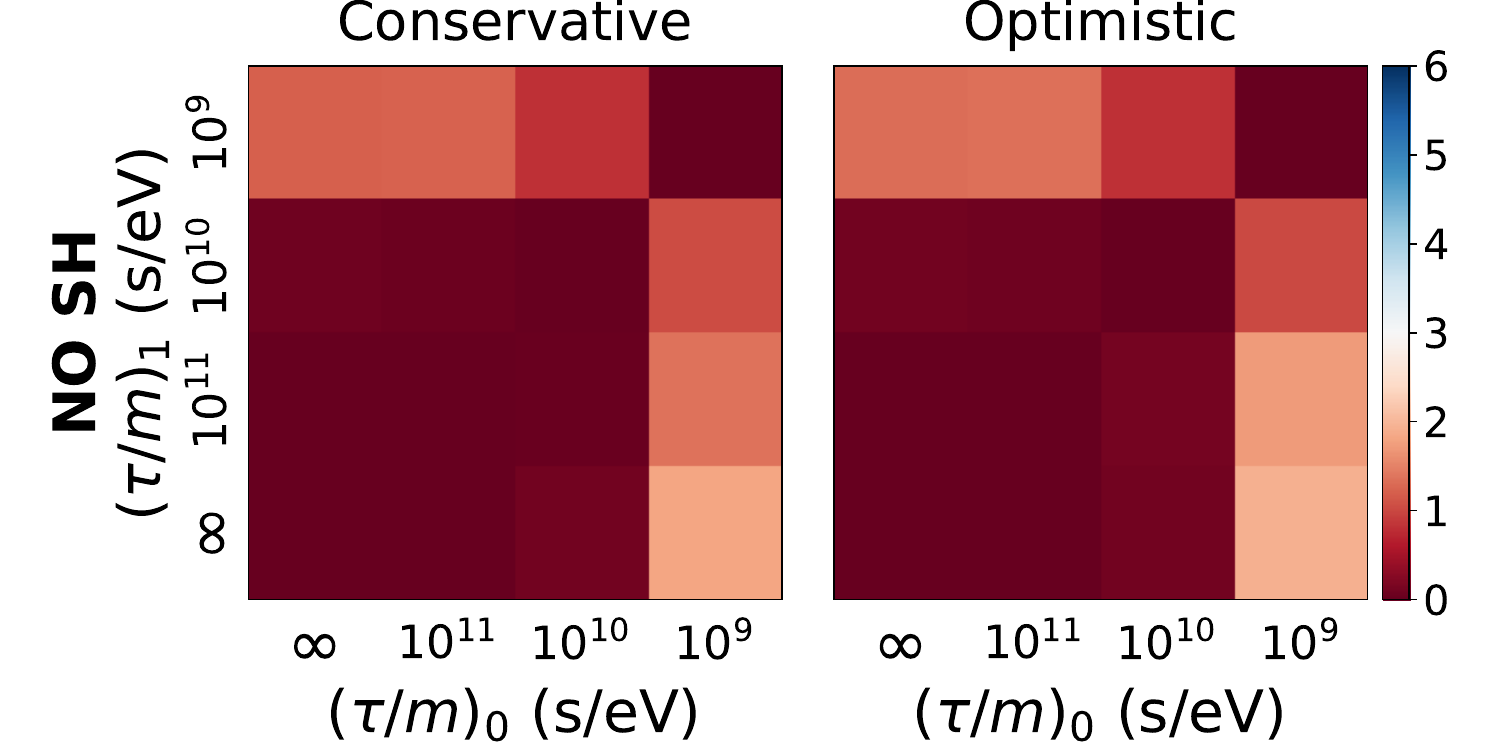}
    \caption{Mean Bayes factor \(\langle\log B_{10}\rangle\) in the combined analysis of the IBD channels in SK-Gd, JUNO and HK, the \(\nu_e\,\)-\(\ce{^40Ar}\) channel in DUNE and including $\nu$-proton scattering in JUNO with optimistic DSNB fluxes.  The results correspond to an idealized case where no background is included for $\nu$-proton scattering. The results correspond to the conservative (left, 40 $\%$ uncertainty on the signal, \(20\%\) on the background) and optimistic (right, 20 $\%$ uncertainty on the signal, \(10\%\) on the background) cases.} 
    \label{fig:BayNOSH}
\end{figure}

Concerning the case NO SH and $\nu$-proton, as mentioned before in relation with fig.\ref{fig:JUNO_nu-p_decay_event_rates}, much larger DSNB rates in this channel could potentially break some of the degeneracies.
To this aim we employed supernova simulations that are on the optimistic side. Indeed we used for the supernova neutrino yields (with no progenitor dependence) Fermi-Dirac distributions at the neutrinosphere, with temperatures 4.5 MeV, 5.5 MeV and 6 MeV for the $\nu_e$, $\bar{\nu}_e$ and $\nu_x$ respectively (and zero chemical potential), including the MSW effect, while using the same evolving core-collapse supernova rate. Figure \ref{fig:BayNOSH} shows the combined analysis of IBD, $\nu_e$-$^{40}$Ar and $\nu$-proton scattering including backgrounds (except for the $\nu$-p channel) and uncertainties on the fluxes and the backgrounds. One can see that one can reach almost strong evidence for $({\tau/m})_0=10^9 $\(\,\si{\second\per\electronvolt}\) (in this idealized case).
However this would require extremely low background in this channel which is currently not expected.

\begin{figure}[htbp]
    \centering
    \begingroup
    \begin{subfigure}{0.7\textwidth}
        \centering
        \includegraphics[scale=0.4]{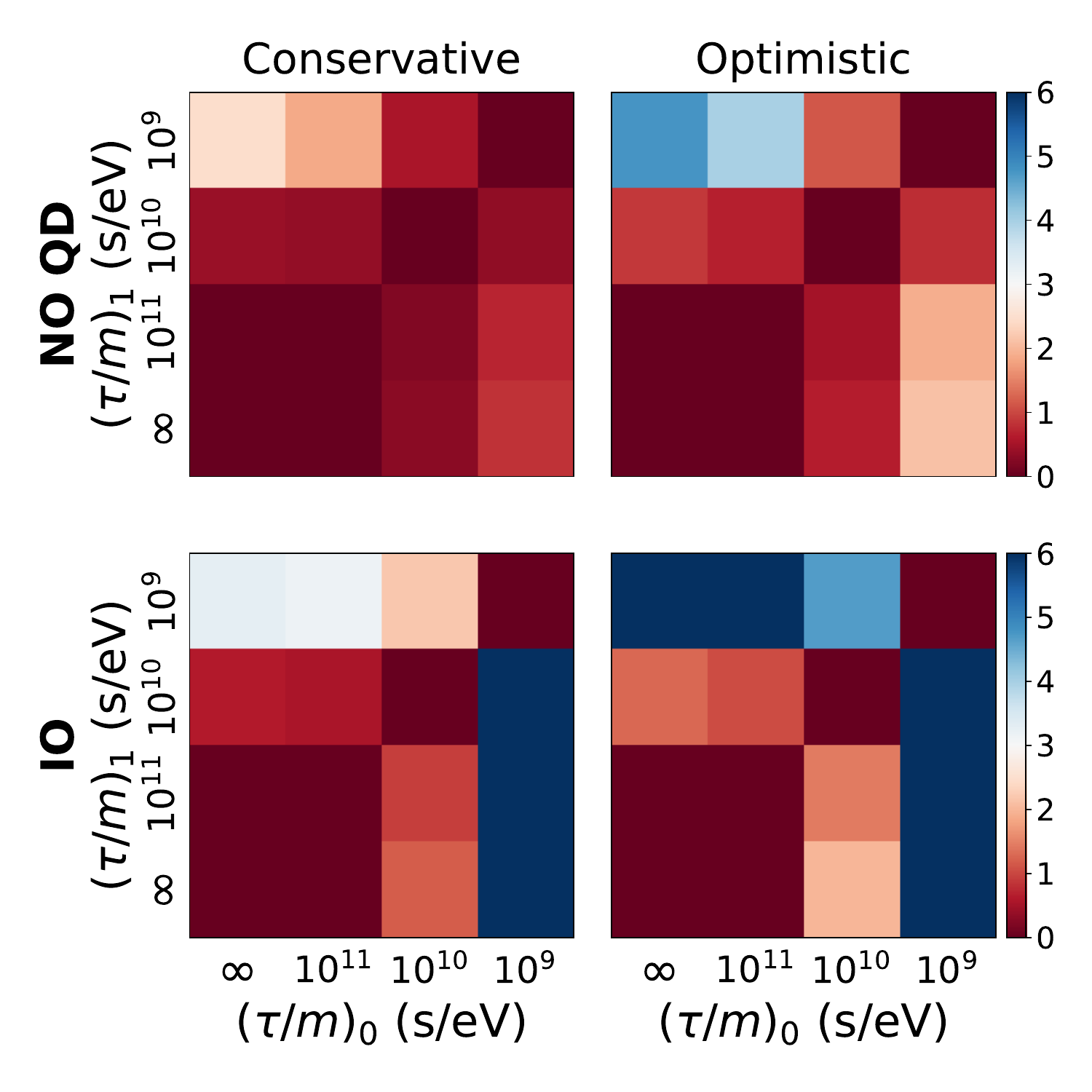}
        \caption{Garching with \(f_\text{BH}=0.41\)}
        \label{fig:bayes_factors_combined_constant_no_fBH041_HK}
    \end{subfigure}
    \hfill
    \begin{subfigure}{0.7\textwidth}
        \centering
        \includegraphics[scale=0.4]{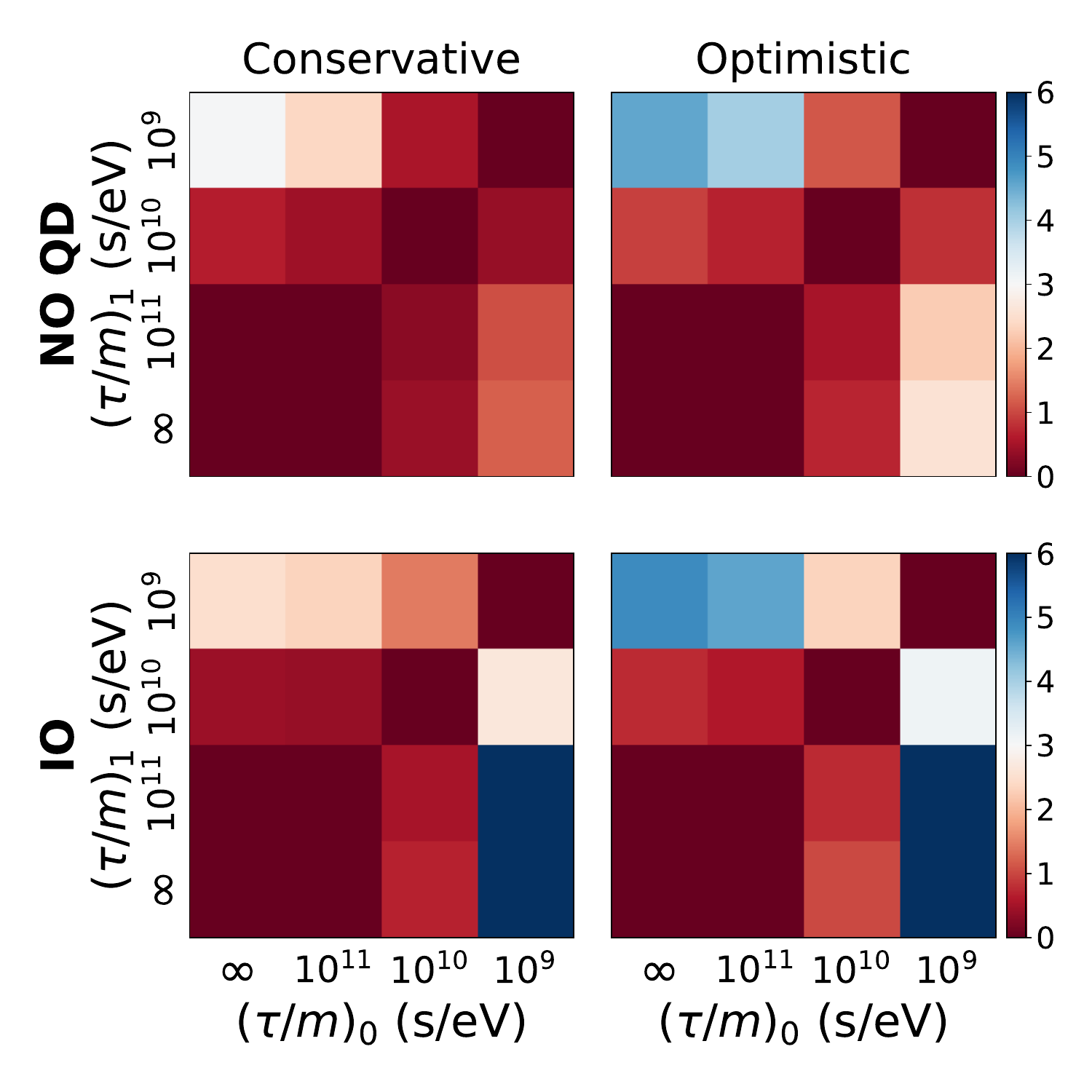}
        \caption{Nakazato with \(Z=0.004\) and \(t_\text{rev}=\SI{100}{\milli\second}\)}
        \label{fig:bayes_factors_combined_constant_no_z0_t1_HK}
    \end{subfigure}
    \endgroup
    \caption{Mean Bayes factor \(\langle\log B_{10}\rangle\) in the combined analysis of the IBD channels in SK-Gd, JUNO and HK and the \(\nu_e\,\)-\(\ce{^40Ar}\) channel in DUNE for the conservative (left, 40 $\%$ uncertainty on the signal, \(20\%\) on the background) and the optimistic (right, 20 $\%$ uncertainty on the signal, \(10\%\) on the background) scenarios in (a) the most optimistic and (b) the most pessimistic simulation.}
    \label{fig:bayes_factors_unc_vs_no_unc_comparison}
\end{figure}

Finally we show how the choice of the supernova simulations impacts the Bayes factors (figure~\ref{fig:bayes_factors_unc_vs_no_unc_comparison}). One can see that \((\tau/m)_0=10^9\,\si{\second\per\electronvolt}\) could be rejected against the case of no-decay with very strong evidence for inverted ordering in the conservative scenario, even in the most pessimistic supernova simulation. Also note that except for NO QD with conservative uncertainties and for $({\tau/m})_1=10^9 $\(\,\si{\second\per\electronvolt}\), the Garching simulation generally yields higher Bayes factors than Nakazato's ones. 

\section{Conclusions and perspectives} \label{sec:conclusions}

In this work we examined for the first time the possibility of untangling the degeneracies between the DSNB predictions in absence and in presence of neutrino non-radiative two-body decay, through a Bayesian analysis. To this aim we employed a \(3\nu\) framework, taking into account the progenitor dependence of the DSNB flux via two different sets of supernova simulations and different possible black hole fractions.  In particular, the latter is the subject of debate in the astrophysical community.
For the decay, we considered the extreme case of normal mass ordering with quasi-degenerate and strongly hierarchical mass patterns, and inverted mass ordering.
We assumed democratic branching ratios for the decaying eigenstates, with equal $\tau/m$ to have only one free parameter, and considered the daughter neutrino to be active in essentially all calculations. At the end we presented results for the invisible decay as well.

First we presented the (integrated) fluxes and event rates for both stable and decaying neutrinos, which show important variations even across the small subset of simulations considered in the present work, that comprises one-dimensional simulations by the Nakazato and the Garching groups. Our predictions are on the conservative side, compared to those in the available literature. 

We presented the results of the Bayesian analysis for experiments first taken individually, and then combined. For the latter, 
we implemented inverse beta-decay in SK-Gd, HK and JUNO, along with the charged-current interaction of \(\nu_e\) with argon in DUNE. 
Moreover we considered also neutrino-electron elastic scattering, charged-current $\nu_e$- and \(\bar\nu_e\)-oxygen interactions in HK, as well as neutrino-proton elastic scattering in JUNO. 
{We find that for these channels the statistics is low and does not significantly improve the Bayesian analysis we performed.  As such, for $\nu$-p scattering our findings are in contrast with previous ones. However, we would like to point out that the behaviors of $\nu$-p DSNB event rates with decay, for NO-SH, is at variance with those for IBD and $\nu_e$-argon scattering.
Therefore, this detection channel could potentially help to break degeneracies between the decay and the no-decay case. To this aim we presented Bayes factors combining IBD, $\nu_e$-argon and $\nu$-proton scattering using optimistic DSNB fluxes and including uncertainties on the fluxes and the backgrounds, while assuming the backgrounds in particular from radioactivity for  $\nu$-proton to be very low. In this idealized case, almost strong evidence against the short lifetime-to-mass ratio could be achieved.

An important outcome of the present work is that, if nature has chosen a normal neutrino mass ordering and a strongly hierarchical mass pattern, then a Bayesian analysis has no discriminating power between no-decay and decay in the range $\tau/m \in [10^9, 10^{11}]$\(\,\si{\second\per\electronvolt}\) even combining all channels in the four experiments.
In normal ordering, while none of the experiments can be expected to discriminate with strong evidence neutrinos decaying with $\tau/m\gtrsim10^9 $\(\,\si{\second\per\electronvolt}\) from stable neutrinos, results are more promising in the quasi-degenerate pattern. In HK-Gd, one could even expect in this case strong evidence against stable neutrinos if $\tau/m \lesssim 10^9 $\(\,\si{\second\per\electronvolt}\), assuming the optimistic scenario for uncertainties. 
For inverted neutrino mass ordering, we found strong evidence against $\tau/m \lesssim 10^9 $\(\,\si{\second\per\electronvolt}\) could be obtained in DUNE and HK, while very strong evidence is expected in JUNO and HK-Gd if neutrinos are stable, even when assuming conservative uncertainties. 

Results are of course enhanced when combining the likelihoods of the different experiments. However, if neutrino masses follow normal ordering and are strongly hierarchical, combination does not suffice to break the DSNB flux degeneracies.  If the ordering is normal, but masses are quasi-degenerate, then stable neutrinos and neutrinos with 
$\tau/m \gtrsim10^{11} $\(\,\si{\second\per\electronvolt}\) could be rejected with strong evidence if $\tau/m \lesssim 10^{9} $\(\,\si{\second\per\electronvolt}\), assuming optimistic uncertainties. In case of inverted ordering, \(\tau/m\lesssim10^9\,\si{\second\per\electronvolt}\) could be rejected with very strong evidence if neutrinos are stable or decay with $\tau/m \gtrsim 10^{10}$\(\,\si{\second\per\electronvolt}\). 
With current-level estimates of the IBD background, the impact of an upgrade from HK to HK-Gd on the combined analysis would remain limited.

We have treated a completely idealized case of vanishing uncertainties and backgrounds to assess the extent to which results could possibly be improved. Reaching such conditions would imply much enhanced discriminating power, even compared to our optimistic scenario, opening the possibility of discriminating $\tau/m=10^{10}$\(\,\si{\second\per\electronvolt}\) against stable neutrinos with (very) strong evidence in normal ordering with quasi-degenerate masses (inverted ordering). Furthermore, the dependence of the DSNB flux on the set of progenitor simulations is reflected in the Bayes factors, the optimistic ones generally yielding higher discriminating power.

In conclusion, we found that current-level uncertainties would suffice to favor \(\tau/m \lesssim 10^{9}\,\si{\second\per\electronvolt}\) against stable neutrinos for normal ordering and quasi-degenerate masses. Whereas if nature has opted for an inverted neutrino mass ordering, then very strong evidence discriminating $\tau/m \lesssim 10^{9}$\(\,\si{\second\per\electronvolt}\) against stable neutrinos could be reached. In contrast, we found that if current hints for normal ordering at $2.5\,\sigma$~\cite{Capozzi_2021}, and for strongly hierarchical mass pattern~\cite{rpp_2024}, turn out to represent the correct neutrino mass pattern, then much-improved knowledge of the DSNB signal and background will not be sufficient to break the DSNB flux degeneracies in upcoming experiments even when  statistics is increased due to optimistic DSNB fluxes, as we have verified numerically.

The constraints  on neutrino non-radiative decay obtained in our analyses, that could be obtained from the DSNB, are competitive with the current tightest bounds that come from cosmology refs.~\cite{Barenboim_2021,Chen_2022}. More precisely, if neutrino masses follow normal ordering with a quasi-degenerate pattern or inverted ordering, then DSNB constraints can be competitive, as long as the mass of the heaviest neutrino obeys \(m_h\gtrsim \SI{0.05}{\electronvolt}\). Indeed, the latest results from ref.~\cite{Chen_2022} indicate an upper bound on neutrino lifetime of \(\tau\gtrsim(2\rightarrow6)\times10^7 \si{\second}\) when considering a \(3\nu\) framework and the assumption of similar couplings among the different eigenstates. 
Furthermore, cosmological observations bring important (albeit model-dependent) bounds on the neutrino masses indicating, in particular that the neutrino mass pattern is likely not quasi-degenerate, and that the bound on the sum of the neutrino masses should be at $\sum m_{\nu} < \SI{0.54}{\electronvolt}$ (CMB alone, 95 $\%$ C.L.) or lower (depending on the priors)~\cite{rpp_2024}. 

Neutrino decay constitute one of the telling examples of the importance of crossing information between particle physics, astrophysics and cosmology to push the frontiers of our knowledge.
Clearly, since the impact of neutrino non-radiative two-body decay on the DSNB differs drastically between normal and inverted mass ordering, determining the neutrino mass ordering through the JUNO, DUNE and HK experiments will be crucial. If future experiments identify the mass ordering to be normal, and the mass pattern to be quasi-degenerate, then breaking degeneracies between DSNB events with neutrinos that decay or do not decay non-radiatively will be possible. But if the mass pattern is strongly hierarchical, the present work clearly shows that more avenues are needed.

\vspace{.5cm}
During the writing of this manuscript Ref. \cite{MacDonald_2024vtw}, which presents a likelihood analysis for invisible neutrino decay in relation with the DNSB, appeared.

\acknowledgments

M. C. Volpe wishes to thank A. Santos for useful discussions concerning the backgrounds in the SK-Gd and HK experiments. This work was supported by the CNRS nucléaire et particules through the NUFRONT Masterproject.  N. Roux acknowledges financial support by the Werner Siemens Foundation. The authors thank the Galileo Galilei Institute for Theoretical Physics for the hospitality and the INFN for partial support during the completion of this work.

\bibliographystyle{JHEP}
\bibliography{DSNB_bayesian_biblio}
\end{document}